\documentclass{aa}
\usepackage{epsf}
\usepackage{natbib}
\bibpunct{(}{)}{;}{a}{}{,}

\def\today{\ifcase\month\or
 January\or February\or March\or April\or May\or June\or
 July\or August\or September\or October\or November\or
 December\fi\space\number\day, \number\year}

\def\todmy{\number\day\space\ifcase\month\or
 January\or February\or March\or April\or May\or June\or
 July\or August\or September\or October\or November\or
 December\fi\space\number\year}

\newcommand{\bdisp} {\begin{displaymath}}
\newcommand{\edisp} {\end{displaymath}}
\newcommand{\beqn} {\begin{equation}}
\newcommand{\eeqn} {\end{equation}}
\newcommand{\beqr} {\begin{array}}
\newcommand{\eeqr} {\end{array}}

\newcommand{\kms}{$\,\mbox{km}\ \mbox{s}^{-1}$}

\newcommand{\HI}{H{\sc i}\ }
\newcommand{\MLstar}{M/L_{\star}}

\begin{document}

\title{High-resolution rotation curves of Low Surface
Brightness galaxies\thanks{based on observations at the
Observatoire de Haute Provence}}
\author{W.J.G. de Blok\inst{1} \and A. Bosma\inst{2}}
\institute{ATNF, CSIRO, P.O. Box 76, Epping NSW 1710, Australia \and
         Observatoire de Marseille,
        2 Place Le Verrier,
        F-13248 Marseille Cedex 4, France}

\offprints{edeblok@atnf.csiro.au}
\date{Received 01 Jan 2000 / Accepted 01 Jan 2001}

\abstract{We present high-resolution rotation curves of a sample of 26 
low surface brightness galaxies. From these curves we derive mass
distributions using a variety of assumptions for the stellar
mass-to-light ratio.  We show that the predictions of current Cold
Dark Matter models for the density profiles of dark matter halos are
inconsistent with the observed curves. The latter indicate a
core-dominated structure, rather than the theoretically preferred
cuspy structure.}

\authorrunning{de Blok \& Bosma}
\titlerunning{LSB galaxy rotation curves}

\maketitle


%\newpage
\section{Introduction}
\label{sec:intro}
\indent

Despite much effort, it is still unclear to what extent rotation
curves can give clues about the distribution of the visible and dark
matter in bright spiral galaxies \citep[e.g.][]{B99,Sellwood}.  It is
thought, however, that low surface brightness (LSB) galaxies and dwarf
galaxies are dark matter dominated, and that therefore the analysis of
their rotation curves can yield directly important information about
the properties and distribution of their associated dark matter halos
\citep{Blok97,Verheijen,Sw1999}.

This implication has far-reaching consequences, as early results by
e.g. \citet{Moore} already showed.  Predictions from cold dark
matter (CDM) simulations were found to disagree with observations of
rotation curves of several dwarf galaxies; the data indicating much
less cuspy distributions of matter than the simulations. It was
thought at the time that this problem could be solved once the effect
of feedback due to star formation was understood. However, the
behaviour of the rotation curves of low surface brightness (LSB)
galaxies is rather similar, and their low star formation rate at
present and in the past indicate that star formation in such galaxies
might never have been important enough to modify their structure
drastically.

Early work on LSB galaxies used 21-cm \HI line work to determine the
rotation curves \citep[e.g.][]{Hulst,Blok96}, and as such the results
are likely to suffer from modest angular resolution effects (commonly
called beam smearing).  Even though these can be partly modelled,
experience shows that direct measurements are preferable (cf.\
discussions in \citealt{B78,KGB}); in particular, supplementary data
in the optical emission lines, such as H$\alpha$ and [N{\sc ii}] are
always useful
\citep[cf.][]{vdK+B, Rub89, Corradi, Sw2000}. Thus a whole industry
has sprung up to combine optical and 21-cm line rotation curves of all
sorts of gas rich galaxies.

For the particular problem of the dark matter distribution in LSB
galaxies \citet*{Sw2000} presented supplementary H$\alpha$ data for five
LSB galaxies already observed in \HI by \citet*{Blok96}, and
concluded that the influence of beam smearing on the \HI curves was
severe enough to question earlier conclusions regarding dark matter
content and rotation curve shape of LSB galaxies. 
\citet{MRdB} and  \citet*{dBMR} (hereafter dBMR) reanalysed these data,
and concluded that the discrepancy between H$\alpha$ and \HI data is
only really significant for one of these five galaxies. Conclusions
regarding the dark matter content and the shape of the dark matter
distribution in LSB galaxies are thus not affected.

In a comparison of pseudo-isothermal and CDM halo models using high-resolution
rotation curves of a sample of a further 29 LSB galaxies dBMR show
that the so-called ``universal'' CDM halo-profile as parameterised in
\citet*{NFW-96} is not a good description of the data: the
rotation curves generally show linear solid-body rise in the inner
parts, which is inconsistent with the steeper rise necessary for the
CDM rotation curves. Rather, the rotation curves prefer a
pseudo-isothermal (i.e.\ core-dominated) halo model. \citet*{dBMBR}
furthermore showed that \emph{all} rotation curves of LSB galaxies
measured so far are consistent with a pseudo-isothermal core
model. Analyses that show that these rotation curves confirm the CDM
NFW halo model \citep[e.g.][]{BS2001} can be traced back to the fact
that at lower resolutions the NFW and pseudo-isothermal models look
sufficiently similar and the errors are large enough that a NFW model
can be made to fit the data.  The higher-resolution data presented in
\citet*{dBMBR} now seem to have settled the observational side of the
debate in favour of core-dominated LSB galaxy halos.  These results
are independent of the stellar mass-to-light ratios one assumes in
constructing these mass models.

It would therefore seem that current models of structure formation and
galaxy evolution need to take into account the fact that most
late-type galaxies have a constant density dark matter \emph{core}
rather than a cusp (see also results by \citealt{Bor,Sal1,Sal2} for
HSB galaxies).

As the signature of the core is clearest at small radii it is
important to find galaxies where these inner parts are well-resolved
and well-sampled.  In this paper we thus supplement the collection of
high-resolution rotation curves of LSB galaxies from \citet*{dBMR}
with curves for an additional 26 galaxies, of which 12 are entirely
new and 14 were already used in the analysis by
\citet*{dBMBR}. The 12 new galaxies in our sample have been
specifically chosen to have small distances so that we can easily
verify the discrepancies between the NFW model and the observations.

This paper is organised as follows: we describe our data in Section 2,
and present the rotation curves in Section 3. Section 4 describes
individual galaxies. In Section 5 we describe the derivation of the
final rotation curves.  The mass models are then presented in Section
6. In Section 7 we discuss these models. Section 8 digresses into the
consequences that systematic observational effects may or may not
have on the data. In Section 9 we discuss the mass densities profiles
inferred from the rotation curves. Section 10 summarises the paper.

\section{Observations and Data reduction}

\subsection{Sample selection}

We constructed a sample of galaxies to observe in H$\alpha$ from lists
of LSB galaxies already observed in the 21-cm \HI line by
\citet*{Hulst} and
\citet*{Blok96}. A representative sample of nearby dwarf galaxies from
the theses of \citet*{Sw1999} and \citet*{Stil} was also included (see
also \citealt{Sw2002opt,Sw2002hi}). The sample was supplemented by two
well-known nearby dwarf galaxies (NGC 1560 and NGC 100) and a galaxy
(UGC 711) from the Flat Galaxy Catalog
\citep{fgc}.  \HI velocity fields were  available for all galaxies 
(except U711 and N100).
Since we observed with a long-slit spectrograph, we determined for
each galaxy the position angle of the major axis of the velocity
field, using the position angle of the major axis of the light
distribution as an additional check.  For U711 and N100 we only used
the position angle of the light distribution: since these are edge-on
galaxies these angles are well determined.

Table 1 lists parameters of these galaxies for which we managed to
obtain useful, high S/N data that will be used in our mass modelling.
The last column in Table 1 gives references to the data presented in
Table 1 not derived in this paper. Images and surface brightness
profiles can also be found in these references. See also Sect.~2.4. 
Table 2 gives a brief
summary of those galaxies that we did observe, but which for various
reasons were deemed to be of insufficient quality to warrant mass
modelling. Our main sample therefore consists of the galaxies listed
in Table 1.

\begin{table*}
\caption{Properties of sample galaxies}
\label{tab:prop}
\begin{center}
\begin{tabular}{llrlllcrrrrrcll}
\hline
UGC &Other & $D$ & $M_R$ & $\mu_{0,R}$ & $h_R$ & $i$ & $R_{\rm max}$ & $V_{\rm max}$ & $V_{\rm sys}$ & $V_{\rm sys,HI}$& PA & bin &Obs&Refs \\
%&&&&\cr
%   &   &(Mpc)&(mag)&(mag$''^{-2}$)&(kpc)&($^{\circ}$)&(kpc)&(km s$^{-1}$)&(km s$^{-1}$)&($^{\circ}$)&($''$)& & \\
(1)&(2)&(3)&(4)&(5)&(6)&(7)&(8)&(9)&(10)&(11)&(12)&(13)&(14)&(15)\\
\hline
%name   & dist & M_R/B    &   & h    & inc&   &  &
--- & F563-1 & 45   & --18.  & 22.6 & 3.5 & 25 & 17.5 & 114 & 3495 &3492 & 161& 5.4 &Jan00&(1)(2)\\
U231 & N100 & 11.2 & --17.7\rlap{$^a$} & -- & --  & 89 &  8.3 & 97  & 841  &--- & 236 & 1.8&Jan00 &---     \\ 
U628   &--- &  65  &--19.2   & 22.1 & 4.7 & 56 & 13.8 & 142 & 5451 & ---& 139 & 1.8&Jan00 &(2)(4)\\
U711   & ---& 26.4 & --17.7\rlap{$^a$} & -- &--& 90 & 15.4 & 92     & 1984& ---& 118 & 1.8&Jan00 &--- \\
U731   &DDO 9 &  8.0 & --16.6   & 23.0 & 1.7 & 57 &  7.0 & 75  &  637 & 638 &  257 & 1.8&Jan00 &(5)\\
U1230  & ---&  51  & --19.1   & 22.6 & 4.5 & 22 & 34.7 & 103 & 3837 & 3835& 112 & 1.8&Jan00 &(2)(4)\\
U1281  & ---& 5.5  & --16.2   & 22.7 & 1.7 & 90 &  5.2 & 57  & 157 & 157   & 218 & 1.8&Jan00 &(5)\\
U3060 & N1560  & 3.0  & --15.9\rlap{$^a$} &23.2\rlap{$^a$}&1.3\rlap{$^a$}&82&  8.3 & 78  & --36 &--36 & 203& 1.8&Jan00 &(3) \\
U3137  & ---& 18.4 & --18.7   & 23.2 & 2.0 & 90 & 30.9 & 100 & 982 & 993  & 255 & 1.8&Jan00 &(5)\\
U3371  &DDO 39 & 12.8 & --17.7   & 23.3 & 3.1 & 49 & 10.3 & 86  & 819 & 818  & 133 & 5.4&Jan00 &(5)\\
U4173  & ---& 16.8 & --17.8   & 24.3 & 4.5 & 40 & 12.2 & 57  & 861  & 865 & 168 &5.4&Jan00 &(5)\\
U4325  &N2552 & 10.1 & --18.1   & 21.6 & 1.6 & 41 &  4.6 & 123 &  523 & 523 & 231 & 1.8&Jan00 &(5)\\
U5005  & ---&  52  & --18.6   & 22.9 & 4.4 & 41 & 27.7 & 99  & 3830 & 3844& 226 &5.4&Jan00 &(2)(4)\\
U5750  & ---&  56  & --19.5   & 22.6 & 5.6 & 64 & 21.8 & 79  & 4169 & 4168& 167 & 1.8&Jan00 &(2)(4)\\
\hline
U3851 & N2366&  3.4&--16.9   &22.6&1.5&59& 5.4 & 55 &102 & 104&42  &1.8&Feb01&(5)\\
%      & N2366&     &&&&& & &&&&\\
U3974&  DDO47&4\rlap{$^b$}    &--14.9\rlap{$^a$}&--  &1.0&30& 3.2 & 67 &282 & 274 &319 &5.4&Feb01&(6)\\
U4278&  I2233& 10.5&--17.7   &22.5&2.3&90& 7.4 & 93 &556 & 559 &173 & 1.8&Feb01&(5)\\
U4426&  DDO52&5.3  &--13.8\rlap{$^a$}&--  &0.6&60& 3.1 & 50 &382\rlap{:} & 395 &185&1.8&Feb01&(6)\\
U5272&  DDO64&6.1  &--14.7\rlap{$^a$}&--  &1.2&60& 2.7& 47 &520 & 525 &97  &1.8&Feb01&(6)\\
U5721&  N3274&6.7  &--16.7   &20.2&0.5&61& 7.3 &79 &542 & 542 &279 &1.8&Feb01&(5)\\
U7524&  N4395&3.5  &--18.1   &22.2&2.3&46& 7.9 &83 &319 & 320 &327 &1.8&Feb01&(5)\\
U7603&  N4455&6.8  &--16.9   &20.8&0.7&78& 5.9& 64&655 & 644  &197 &1.8&Feb01&(5)\\
U8286&  N5023&4.8  &--17.2   &20.9&0.8&90& 5.9 &84 &403& 407& 208 &1.8&Feb01&(5)\\
U8837&  DDO185&5.1 &--15.7   &23.2&1.2&80& 2.1 & 50 &148& 135 & 22  &1.8&Feb01&(5)\\
U9211&  DDO189&12.6&--16.2   &22.6&1.2&44& 8.3 & 64 &685& 686 & 287 &5.4&Feb01&(5)\\
U10310& Arp 2&15.6 &--17.9   &22.0&1.9&34& 9.0& 75&720 & 718 &202 &1.8&Feb01&(5)\\

\hline
\end{tabular}
\medskip\noindent
\begin{minipage}{177mm}

\noindent
Explanation of columns: {\bf (1)} UGC number {\bf (2)} Other
identification {\bf (3)} Distance [Mpc] (see text) {\bf (4)} Absolute
magnitude in $R$-band corrected for Galactic extinction [mag]. 
{\bf (5)} Central surface brightness in
$R$-band, corrected for Galactic extinction and inclination [mag
arcsec$^{-2}$] {\bf (6)} Scale length [kpc] {\bf (7)} Inclination
[$^{\circ}$] {\bf (8)} Maximum radius rotation curve [kpc] {\bf (9)}
Maximum rotation velocity [\kms] {\bf (10)} Systemic velocity [\kms]
{\bf (11)} Systemic velocity derived from \HI data [\kms] {\bf (12)}
Position angle slit [$^{\circ}$] {\bf (13)} Binning interval rotation
curve [arcsec] {\bf (14)} Observing run {\bf (15)} References: (1)
\citet*{Blok96} (2) \citet*{Blok95} (3) \citet*{Broeils} (4)
\citet*{Hulst} (5) \citet*{Sw1999} (6) \citet*{Stil}. {\bf Notes:}
$(a)$ $B$-band data. $(b)$: distance from \citet*{Walter}

\end{minipage}
\end{center}
\end{table*}

\begin{table}
%\begin{center}
\caption{Galaxies not analysed\label{tab:other}}
\begin{tabular}{ll}
\hline
Name & Reason for rejection\\
\hline
\noalign{\smallskip}
\noalign{\bf Jan 2000 observations}
\noalign{\smallskip}
F561-1& low S/N\\
F564-V3 & no H$\alpha$ detected\\
F568-6& only bulge visible\\
U5999 & suspect slit position\\
DDO154& low S/N, suspect slit position\\
DDO127& low S/N\\
\hline
\noalign{\smallskip}
\noalign{\bf Feb 2001 observations}
\noalign{\smallskip}
N4214 & suspect slit position\\
DDO125& low S/N\\
DDO43& low S/N\\
DDO168& low S/N\\
\hline
\end{tabular}
%\end{center}
\end{table}

\subsection{Optical spectra}
\label{sec:observ}

The emission line observations have been made with the 193-cm
telescope at the Observatoire de Haute Provence and its long-slit
Carelec spectrograph from 2-9 Jan 2000 and 22-27 Feb 2001. The
spectrograph has been described in detail by \citet*{Lemaitre}.  As a
detector, we used a EEV 2048 x 1024 CCD chip. The slit length on the
sky was about 5.5\arcmin, and we used a 2.0\arcsec\ slit-width. This
works out to a pixel size of 0.6\arcsec\ $\times$ 45.7 km
s$^{-1}$. Spectral resolution was 54 km s$^{-1}$ FWHM. Typical
exposure times were 1 hour per spectrum, preceded and followed by
calibration spectra from a Neon arc lamp. The typical seeing during
the observations was $\sim 2\arcsec$.  Care was taken to align the
slit with the optical centre of the galaxy.  This was done with
off-set pointing from nearby stars. In many cases the central part of
 the galaxy was visible in the guiding camera, and the accuracy of the
off-set procedure could be verified. For a few of these galaxies we
repeated the off-set procedure a few times, and found that the galaxy
acquisition was indeed repeatable, with an error of less than
1$''$. See Sect.~8 for a full discussion of the effects of
(mis-)pointing.
   
\subsection{Data reduction}

The data were corrected for instrumental effects (bias, flat-field,
etc.) in the standard manner. The spectra were then further reduced
with the Figaro package: the calibration spectra were averaged, and
for each of the spectra a wavelength solution was found and the
wavelength calibration applied. Cosmic rays were corrected with an
automated method, which was verified interactively.

The data were then further processed using the {\sc emsao} programme
of the {\sc rvsao} package \citep{emsao} within the {\sc iraf}
environment.  Emission lines were traced and their wavelengths determined
using known night sky lines. Spectral shifts were then converted into
barycentric radial velocities. The spectra were averaged every 3 pixels
in spatial direction (1.8\arcsec). Some spectra with low S/N were
averaged every 9 pixels (5.4\arcsec) as indicated in Table~1.

\subsection{Data collected from the literature}

Since our aim is to make mass models of our galaxies, we collected
supplementary data from the literature, basically optical surface
photometry and \HI data. The references are listed in the notes to
Table~1 (see also \citealt{Sw2002opt,Sw2002hi}).  For 22 out of the 26
galaxies studied here optical $R$-band ($B$-band where noted in
Table~1) is available, as well as \HI imaging and \HI and H$\alpha$
rotation curves. Of these 22 galaxies, 5 are (close to) edge-on and
because of the uncertainties involved in de-projecting surface
brightness profiles of edge-on galaxies we will not consider their surface
photometry here. See also Sect.~6.1.

For U628 we only have $R$-band photometry and H$\alpha$ velocity data.
For two galaxies (U711 and N100) we only have H$\alpha$ data. For
U10310 we only have \HI and H$\alpha$ data.  For these four galaxies
we can construct only limited mass models.

Generally we adopted the distances given in the source papers
(corrected to $H_0 = 75$ km s$^{-1}$ Mpc$^{-1}$), except for DDO 47
where we used the revised distance given in \citet{Walter}, and for
U711 where we used the recession velocity. Inclinations were also
adopted from the source papers.  These inclinations are for the major
part kinematically derived and as such only have a few degrees error
at most.  It should be emphasised that the shape of the rotation curve
and hence applicability of any particular model does not depend on
inclination, and thus an uncertainty of a few degrees does not affect
our main conclusions.  
The scale lengths and surface brightnesses were
are also taken from the references listed in Table~1.  We refer to
these papers for precise details.  Here it suffices to say that they
were derived using a simple straight-line fit to the exponential part
of the profile (most of the sample galaxies are dominated by an
exponential disk).

\section{Results}

\subsection{Raw HI and H$\alpha$ data}

Our principal results are collected in Figs.~1 to 5.  Fig.~1 shows the
raw velocity curves derived from the H$\alpha$ data for the galaxies
listed in Table~1.  Fig.~2 shows the raw data of the galaxies listed
in Table~2. These latter data will not be analysed further.

\begin{figure*}
\epsfxsize=0.5\hsize
\epsfbox{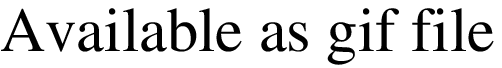}
\caption{Raw H$\alpha$ velocity curves. 
The centre and systemic velocity are indicated by dotted lines.}
\label{panels-01}
\end{figure*}

Fig.~3 presents overlays of the H$\alpha$ data on a major axis
position-velocity diagram from the \HI data. Such a presentation has
the advantage of immediately showing any discrepancies between the two
datasets, if they exist.  The \HI data are taken from the literature,
with references given in Table~1. Also indicated is the \HI rotation
curve as derived from the position-velocity diagram.
These overlays will be discussed in more detail in Sect.~4.

\begin{figure*}
\epsfxsize=0.5\hsize
\epsfbox{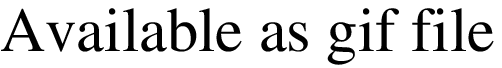}
\caption{Raw H$\alpha$ velocity curves of galaxies not used in modelling.
 The centre and
systemic velocity are indicated by dotted lines.}
\label{panels-01}
\end{figure*}

\begin{figure*}
\begin{center}
\epsfxsize=0.5\hsize
\epsfbox{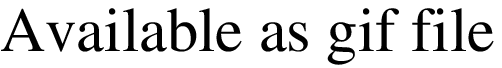}
\caption{Overlay of the raw H$\alpha$ data on \HI position velocity diagrams. 
See Table~1 for the source of the \HI data. The H$\alpha$ data are
indicated by the black dots. The rotation curves based on \HI data are
indicated by a dark-grey line.}
\end{center}
\end{figure*}

\subsection{Symmetrised Rotation curves}

In most cases the continuum emission from the central parts of the
galaxies was visible in the spectra. This was used to determine the
central position in the spectra and the systemic velocities. In a few
cases (DDO185, U10310, U5999) the centre was not seen and we varied
the central position and systemic velocity over a small range until we
found the values that gave the maximum symmetry.  The systemic
velocities based on the H$\alpha$ data are given in Table~1.  A
comparison with the systemic velocities derived from the HI data (also
listed in Table~1 and taken from the source papers) shows a good
correspondence with the H$\alpha$ velocities. The absolute difference
is 5 km s$^{-1}$ or less, except in 6 cases where the difference lies
between 10 and 14 km s$^{-1}$.  In 4 of the 6 cases the HI data were
obtained using a velocity resolution of over 20 km s$^{-1}$, implying
that the comparison with H$\alpha$ is still very favourable.  Only for
DDO 47 and DDO 52 do we find a difference of $\sim 1.5$ HI channel
widths.  Given the limited spatial and velocity resolution of the HI
data it is not clear where this difference originates.

\begin{figure*}
\begin{center}
\epsfxsize=0.5\hsize
\epsfbox{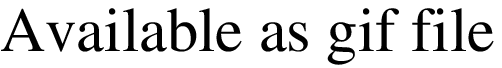}
\caption{Symmetrised and folded rotation curves. Different symbols
indicate different sides of the galaxies. The rotation velocities
have been corrected for inclination.}
\label{folded}
\end{center}
\end{figure*}

Figure~\ref{folded} show the folded and symmetrised rotation curves,
corrected for inclination. 

\begin{figure*}
\begin{center}
\epsfxsize=0.5\hsize
\epsfbox{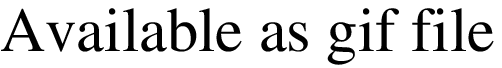}
\caption{Comparison of the folded H$\alpha$ rotation curves (black dots) with
the \HI rotation curve (heavy line).}
\label{compare}
\end{center}
\end{figure*}

Figure~\ref{compare} compares the folded H$\alpha$ rotation curves with the
\HI rotation curves as presented in the literature 
(see Table 1 for references).  It is clear that \emph{grosso modo}
there is reasonable agreement. A small number of galaxies were
obviously affected by beam-smearing, just as for some galaxies the
H$\alpha$ rotation curve shows signs of non-circular motion.

\section{Description of individual galaxies}

In this section we describe the various rotation curves of each galaxy
in some detail. We will compare the \HI and the H$\alpha$ data, as
well as the \HI and H$\alpha$ rotation curves as presented in Figs.~1
and 3-5. For some galaxies we make a comparison with independent data
from the literature.

{\bf F563-1}: The H$\alpha$ data show reasonable agreement with the
\HI data, showing that beamsmearing is not a major problem. At the
receding side the H$\alpha$ data disagree slightly with the \HI
curve. An independent observation by dBMR of this galaxy is available
and agrees with our observation (see Fig.~6); any systematic
differences between the data sets is less than the uncertainties in
individual data points.  Obtaining high-resolution H$\alpha$ rotation
curves of LSB galaxies is thus a repeatable exercise.

{\bf U5005}: The H$\alpha$ data agree with the \HI position-velocity
diagram, but the resulting high-resolution curve rises less steeply
than the \HI curve.  The \HI data have only a low spatial resolution
of $40\arcsec$ \citep{Hulst} and uncertain corrections for beam
smearing are thus unavoidable.

{\bf U1230}: The H$\alpha$ data are of high-quality, and shows that
beamsmearing has affected the $40''$-resolution \HI curve
\citep{Hulst}. The high-resolution curve rises more steeply,
though still with a solid-body signature.

{\bf U5750}: The \HI data are poor, but in agreement with the
H$\alpha$ data. The linearly rising \HI curve also shows reasonable
agreement with the H$\alpha$ data. For this galaxy an independent
observation of dBMR is available and shows excellent agreement
(Fig.~6), again showing that systematic effects due to telescope pointing
etc.\ are negligible.

{\bf U731/DDO9}: the H$\alpha$ is weak at the receding side and
superimposed on a night sky line.  This makes it virtually impossible
to say anything about the H$\alpha$ velocities between $+0.5'$ and
$+2'$.  The H$\alpha$ is stronger at the approaching side, and agrees
well with the ridge in the \HI $pV$ diagram.  The H$\alpha$ curve
rises more steeply than the \HI curve, which turns over too
quickly. U731 is a ``kinematically lopsided'' galaxy
\citep{Swlop1999}, and the \HI rotation curve of the approaching side
is different from that of the receding side.  The H$\alpha$ data
agrees well with the steep approaching side of the
\HI position-velocity diagram.  The H$\alpha$ \emph{curve} is however 
significantly steeper than the \HI \emph{curve}. This discrepancy
cannot be explained by lopsidedness effects, which are much smaller
than the difference shown here.  The \HI curve extends over $\sim 8$
beams in radius, and this galaxy thus demonstrates that even for
well-resolved galaxies corrections for beam-smearing are ambiguous,
and do not always yield the correct outcome.

{\bf U1281, U3137, U3371/DDO39, U4325/N2552}: For these four galaxies
there is excellent agreement between the \HI position velocity data
and the H$\alpha$ curve.  There is some disagreement between the \HI
and H$\alpha$ rotation curves for U4325 and U3371. In these cases the
\HI curves rise more steeply than the H$\alpha$ data. The agreement
between the \emph{raw} \HI and H$\alpha$ data suggests that the cause
of this disagreement is likely an overcorrection for beamsmearing in
the \HI data. While suited for correcting for large-scale resolution
effects, beam smearing corrections thus give non-unique solutions on
small scales --- a rotation curve that looks like it has linear rise,
can in reality be a steeply rising rotation curve (as the
beam-smearing corrections in this case usually assume),
\emph{or} it can be an intrinsically linearly rising curve, as the
H$\alpha$ data in this case show.

{\bf U4173}: the H$\alpha$ data are rather poor in quality, but do
seem to agree with the \HI data.

{\bf N100/U231}: good agreement between the data sets. Given the edge-on
orientation of the galaxy, it reaffirms the transparency of this small
galaxy, cf. \citet{B92}.

{\bf N1560/U3060}: excellent agreement between \HI and H$\alpha$. This curve
is a combination of three spectra covering the entire extent of the
optical disk of N1560. The H$\alpha$ curve shows clear signs of
non-circular motions in the disk of N1560 (high-velocity gas near star
forming regions, etc.).

{\bf N2366/U3851/DDO42}: The H$\alpha$ rotation-curve is dominated at the
approaching side by the effects of non-circular motions. This position
corresponds to a large star forming region \citep{Hunter}, that
is disturbing the dynamics on that side of the galaxy.   The
unaffected 
raw H$\alpha$ data
show excellent agreement with the \HI position-velocity diagram, the
H$\alpha$ rotation curve, however, rises steeper than the \HI curve.

{\bf N3274/U5721, N4395/U7524}: In general there is excellent
agreement 
between the
various raw data sets. The H$\alpha$ curves rise a bit faster than
the \HI curves.

{\bf N4455/U7603:} The H$\alpha$ agrees well with the \HI position velocity
diagram.  The \HI rotation curve seems to over-estimate the rotation
velocity in the outer parts by a significant amount. For this galaxy
we found a $\sim 10$ km s$^{-1}$ difference between the systemic
velocities of the \HI and H$\alpha$ curves. The velocity resolution of
the \HI data is, however, 25 \kms, so this difference may not be significant.

{\bf N5023/U8286:} Excellent agreement between the data sets, except between
$\sim 50\arcsec$ and $\sim 100\arcsec$ where the \HI curve
overestimates the 
velocity.
The good agreement in the inner parts of this edge-on galaxy confirm again 
that  these late-type galaxies are virtually transparent.

{\bf U10310/Arp2:} The H$\alpha$ data of this galaxy is rather poorer
in quality than the previous couple of data sets. Nevertheless the
data sets are in good agreement, except perhaps the \HI rotation curve
which seems to overestimate the true rotation velocity.

{\bf DDO47/U3974:} The H$\alpha$ curve is rather poor in quality, but shows
reasonable agreement with the \HI curve.

{\bf DDO52/U4426:} Rather poor H$\alpha$ data but a reasonable match with the
\HI\ curve, except in the outer parts, where beam smearing effects presumable
cause a small discrepancy.

{\bf DDO64/U5272:} Good H$\alpha$ data show that the \HI curve suffers from
slight beam smearing effects. Non-circular motions presumably cause the
difference between approaching and receding sides at $r\sim 30\arcsec$.

{\bf DDO185/U8837:} The H$\alpha$ data are of good quality, but do not
correspond very well with the \HI position-velocity overlay nor the
\HI curve. The most likely explanation is non-circular motions in
especially the approaching side, where the DSS image shows what is
likely a large star formation complex. The receding side does show a
linear rise.  We do find an H$\alpha$ systemic velocity that differs
$\sim 15$ km s$^{-1}$ from the \HI systemic velocity. Alternatively,
as this is one of the galaxies where the continuum emission of the
central part could not be detected, we may have chosen the wrong
systemic velocity. This curve is therefore rather uncertain.

{\bf DDO189/U9211:} The H$\alpha$ data are of average quality. The various
data sets are all consistent with each other.

In summary the comparison between \HI and H$\alpha$ shows that the \HI
curves in this sample tend to suffer from two systematic effects. The
low-resolution curves clearly suffer from beam-smearing which
underestimates the rotation velocities. This is apparent in e.g.\
U1230. A second effect is apparent for a small number of galaxies
where the \emph{raw} \HI and H$\alpha$ data are an excellent match,
but where the \HI \emph{curve} overestimates the true rotation
velocity. Because of the good match between the raw data sets, this
systematic effect must have been introduced as a result of
beam-smearing corrections; U4325 is a good example.

\begin{figure}
\epsfxsize=0.95\hsize
\epsfbox[18 157 565 690]{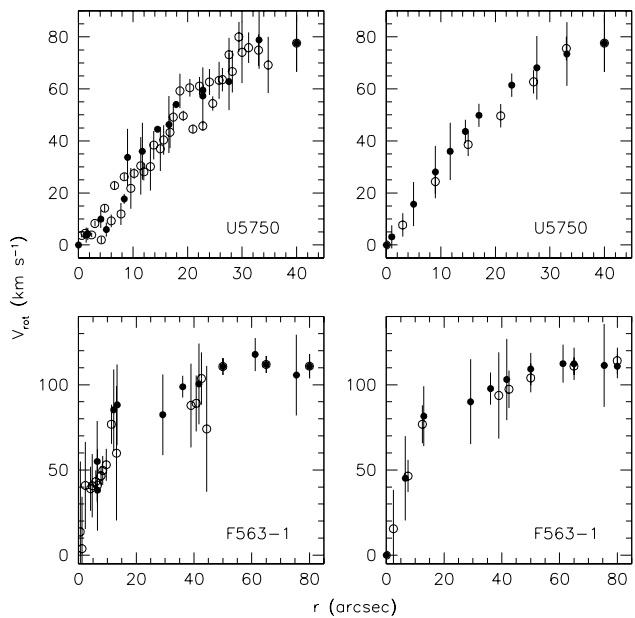}
%\special{psfile=bosmacomp.ps hoffset=-16.0 voffset=-320.0 hscale=45 vscale=45}
%\vspace{9.0truecm}
\caption{Comparison of the data obtained by dBMR
(open circles) and our data (filled circles), for two galaxies in
common. At left is the folded original data, and at right the final
adopted smooth rotation curves. The agreement is reasonable for UGC
5750 and F563-1.}
\label{bosmacomp}
\end{figure}

This comparison underscores the need for this kind of high-resolution
H$\alpha$ data even in moderately well-resolved
galaxies. Beam-smearing corrections are useful in deriving the general
properties of rotation curves. For example, they can tell us what
the general shape of the original curve was, or how likely it is that
a particular rotation curve shape can give rise to the observed
data. When it comes to determining an accurate slope or detailed shape
for an individual galaxy, beam smearing correction are frequently
ambiguous.

\section{Deriving hybrid rotation curves}

In many cases the \HI data extend further out than the H$\alpha$
data. Due to the flattening of the \HI curve towards larger radii,
resolution effects become less important, and to get a more extended
rotation curve it therefore makes sense to combine the \HI and
H$\alpha$ data. We have defined hybrid H{\sc i}/H$\alpha$ curves,
always giving preference to H$\alpha$ data where available. In
practice this meant that we simply adopt the H$\alpha$ rotation
velocities interior to a radius $r_{\rm{H}\alpha}$, usually
corresponding to the outermost point of the H$\alpha$ rotation curve,
and the \HI velocities outside that radius.  Table~\ref{tab:comb}
lists these radii, along with supplementary notes.

\subsection{Asymmetric Drift}
In order to combine the \HI and H$\alpha$ rotation curves one should
in principle correct for the pressure gradients in the gas.  This
correction for asymmetric drift is given by $$v_c^2 = v_{\phi}^2 -
\sigma^{2} \left( \frac{\partial \ln \rho}{\partial \ln R} +
\frac{\partial \ln \sigma^{2}}{\partial \ln R} \right),$$ where $v_c$
is the true rotation velocity, $v_{\phi}$ the observed gas rotation
velocity, $\sigma$ the velocity dispersion in the gas, and $\rho$ the
volume density. The velocity dispersion in late-type galaxies is found
to lie between 7 and 10 km s$^{-1}$ and is to first order constant with
radius. Assuming a constant scale height we then find that the correction
for asymmetric drift is typically $\sim 3$ km s$^{-1}$ or less.
Given other uncertainties due to non-circular motions, inclination, etc. the
asymmetric drift corrections are negligible and we have therefore not applied
this correction.

\begin{table}
\caption{Combining \HI/H$\alpha$\label{tab:comb}}
\begin{tabular}{lrlr}
\hline
Name & $r_{{\rm H}\alpha} (\arcsec)$&Name & $r_{{\rm H}\alpha} (\arcsec)$  \\
\hline
F563-1  & 60     &N1560   & 200\\
U628    &---$^a$&IC2233 &---$^b$\\
U711    &---$^a$&N2366  &200\\
U731    & 40    &N3274  &70\\
U1230   & 90  &DDO64    &80\\
U1281   & 115 &DDO52    &80\\
U3137   & 88  &DDO47    &100\\  
U3371   & 104 &DDO185   &---$^b$\\
U4173   & 10   &DDO189  &60\\
U4325   & ---$^b$&N5023 &180\\
U5005   & 65  &N4455    &100\\
U5750   & 39  &N4395    &250\\
N100    &---$^a$&U10310 &70\\
\hline
\end{tabular}

\begin{minipage}{60mm}
{\bf Notes:} (a) No \HI curve; (b)\HI not used
\end{minipage}
\end{table}

\subsection{Deriving smooth rotation curves}
Some of the main assumptions that are made when deriving mass models
from rotation curves is that there is symmetry, that all mass is on
circular orbits and  there is continuity with radius. The raw data
still show some scatter and non-circular motions, which can
occasionally result in virtual or ambiguous rotation velocities.  In
order to estimate the (smooth) radial run of the potential we want to
represent the data by a curve whose velocity and velocity-derivative
change smoothly, retaining small-scale details. As we are combining
\HI and H$\alpha$ points we also want the density of the different
data points to be approximately equal, in order to prevent the densely
sampled H$\alpha$ curves from dominating the fits. We have used a
local regression method to derive the smooth curve \citep{Loader}. This
method fits polynomial pieces \emph{locally} to subsets of data but
does not impose any \emph{global} functional description (see dBMR for
a more extensive description).  As this process inevitably introduces
a small degree of correlation between adjacent data points, the smooth
(H$\alpha$) curves were resampled every 6$''$.

The final errorbar in these resampled data points consists of two
components: {\bf (i)} the average measurement uncertainty, derived
from the average weighted measurement error in each 6$''$ interval;
{\bf (ii)} an error estimate due to large-scale asymmetries and
non-circular motions. We used the difference between the weighted mean
velocity and the smooth curve in each interval as an estimate.  These
two uncertainties were added quadratically giving a conservative
estimate for the uncertainty.

A general problem with H$\alpha$ rotation curves is that the formal
errorbars of well-sampled, high S/N data points become unrealistically
small (sometimes less than 1 km s$^{-1}$).  These errorbars no longer
have any physical significance but simply indicate that the Gaussian
fit to the profile was well-determined.  A similar problem also exists
with \HI curves where the tilted-ring fitting programs produce error
bars that give the uncertainty in the fit and \emph{not} the actual
uncertainty due to profile width etc.  Realistically, observational
and physical uncertainties make it difficult to determine a meaningful
rotation velocity with an accuracy more than a few km s$^{-1}$.  As
the final step in our derivation of the rotation curves we have
imposed a minimum error of 4 km s$^{-1}$ (prior to inclination
correction) on each data point.

\begin{figure*}
\centering
\epsfxsize=0.5\hsize
\epsfbox{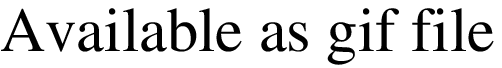}
\caption{Final hybrid smooth rotation curves (full and dashed line) overlaid on
original raw hybrid H$\alpha$\HI data (grey symbols). The smooth
curves are a good representation of the raw data.
\label{smoothcomp}}
\end{figure*}

We compare the resampled smooth curves with the raw data in
Fig.~\ref{smoothcomp}. To avoid cluttering the plot we represent the
smooth curves as a line. The full curves, complete with data points,
are also shown in Fig.~\ref{decomps}. It is evident that the method
does not introduce any systematic effects that affect the shape of the
curve.  For an extensive discussion see dBMR.  The curves shown in
Fig.~\ref{smoothcomp} will be used as the basis for our mass
modelling.

\section{Mass models}

In order to find the signature of the dark matter halo one needs to
decompose the observed rotation curve in a number of separate
dynamical components, described below.

\subsection{Stellar and gas components} For the stellar disk the $R$-band
photometry presented in \citet{Blok95} and \citet{Sw1999} (see also
\citealt{Sw2002opt}) was used.  For NGC 1560 we used the $B$-band
photometry presented in
\citet{Broeils}.  For DDO47 and DDO64 we used the radial surface
brightness profiles from \citet{Stil}. These profiles are presented
as relative to the sky-level without absolute calibration and we
therefore used the surface photometry presented in \citet{Hopp}
and \citet{Makarova} to put the profiles on an absolute $R$-band surface
brightness scale. (Since the Stil photometry is in the $I$-band, this assumes
that there are no strong $R-I$ gradients.)
See Table 1 for more details. 

The rotation curve of
the disk was computed using the {\sc rotmod} task in {\sc gipsy}.  The
disk was assumed to have a vertical sech$^2$ distribution with a scale
height $z_0 = h/6$ \citep{vdK}.  The rotation curves of the
stellar component were resampled at the same radii as the smooth
curves. We assume $\MLstar$ constant with radius, which is a
reasonable assumption for the range of variation expected from
plausible stellar populations \citep{DeJong}.

For the gas component we used the \HI surface densities from the
references given in Table 1. They were scaled by a factor of 1.4 to
take into account the contribution of helium and metals. Their
rotation curve was derived assuming the gas is distributed in a thin
disk.  For a few nearby galaxies we had to slightly smooth the \HI
radial profile, mainly because small-scale structure at small radii
cause unrealistically large fluctuations in the gas rotation curve
and hence the potential. At large radii this small-scale structure is
averaged out so the \HI profile there is a good description of the
global \HI structure.

\subsection{Dark halo} 

After subtraction of the above components from the observed rotation
curve, any residuals are usually taken to indicate the presence of a
dark matter halo.  The halos are usually parametrised by fitting a
model to these residuals. The quality of the fit tells us something
about the applicability of any particular model.  Here we restrict
ourselves to two well-known models: the pseudo-isothermal halo, used
in most of the classic rotation curve studies, and the CDM NFW halo.

\subsubsection{Pseudo-isothermal halo}

We assume a spherical pseudo-isothermal
halo with a density profile
 \begin{equation} \rho_{ISO}(R) = \rho_0 \Bigl[ 1 + \Bigl( {{R}\over{R_C}} \Bigr)^2 \Bigr]^{-1},
 \end{equation}
 where $\rho_0$ is the central density of the halo, and $R_C$ the core
 radius of the halo.  This density profile results in the rotation
 curve
 \begin{equation} 
   V_{\rm halo}(R) = \sqrt{ 4\pi G\rho_0 R_C^2 \Bigl[ 1 -
     {{R_C}\over{R}}\arctan \Bigl( {{R}\over{R_C}} \Bigr) \Bigr] }.
\end{equation}
The asymptotic velocity of the halo, $V_{\infty}$, is given by
\begin{equation} 
  V_{\infty} = \sqrt{ 4 \pi G \rho_0 R_C^2 }.
\end{equation}
To characterise this halo only two out of the three parameters
$(\rho_0, R_C, V_{\infty})$ are needed, as equation (3) determines the
value of the third parameter. 

\subsubsection{NFW halo}

The NFW mass density distribution \citep{NFW-96,NFW-97} takes the form
\begin{equation}
\rho_{NFW}(R) = \frac{\rho_i}{\left(R/R_s\right)
\left(1+ R/R_s\right)^2}
\end{equation}
where $R_s$ is the characteristic radius of the halo and $\rho_i$ is
related to the density of the universe at the time of collapse.  These
parameters are not independent and are set by the cosmology.  They are
usually expressed in a slightly different form by the concentration
parameter $c = R_{200}/R_s$, and the rotation velocity $V_{200}$ at
radius $R_{200}$. The latter is the radius where the density contrast
exceeds 200, roughly the virial radius \citep{NFW-96}.  The
rotation curve of the NFW halo then is given by
\begin{equation}
V(R) = V_{200} \left[\frac{\ln(1+cx)-cx/(1+cx)}
{x[\ln(1+c)-c/(1+c)]}\right]^{1/2},
\end{equation}
where $x = R/R_{200}$.

\subsection{Stellar Mass-to-light ratios}

One of largest uncertainties in any mass model is the precise value of
$\MLstar$. Though broad trends in $\MLstar$ have been measured and
modelled \citep{Bott,Bell}, the precise value from
galaxy to galaxy is unknown, and depends on extinction, star formation
history, initial mass function, etc.  Because of this problem we
present disk-halo decompositions using four different assumptions for
$(M/L)_{\star}$ for those galaxies with photometry available. For the
galaxies without photometry only the first (minimum-disk) model is
presented.

\begin{enumerate} 

\item \emph{minimum disk.} This ignores all visible components and assumes
that the observed rotation curve is entirely due to dark matter. This
gives a strong limit on the amount of dark matter in a galaxy.
This is what is usually called ``minimum disk'' in the CDM analyses.

\item \emph{minimum disk + gas.} This is the classical definition 
of minimum disk mostly used in observational rotation curve papers
(and usually called ``minimum disk'' there). The contribution of the
gas is taken into account, but $\MLstar$ is assumed to be zero.

\item \emph{constant $\MLstar$.} Here $\MLstar$ is fixed to a constant and 
representative value for all galaxies. In this case we have chosen
$\MLstar(R)=1.4$ which is at the ``blue'' or light-weight end of the
range of plausible values derived (under reasonable assumptions for
IMF and star formation history) from population synthesis models using
the observed colours as constraints.  See dBMR for a further motivation
of this value.  By using this ``lightweight'' value we give the 
halo models maximum opportunity to fit the data.

\item \emph{maximum disk.} Here the rotation curve of the stellar
component is scaled to the maximum value allowed by the smooth curve,
without exceeding it \citep{maxdisk86}. In practice the
stellar curve is scaled until the inner points match those of the
smooth curve, but avoiding a hollow halo.
\end{enumerate}

Each of the rotation curves was fitted using the {\sc gipsy} task {\sc
rotmas}. The program determines the best-fitting combination of $R_C$
and $V_{\infty}$ (for the pseudo-isothermal halo) or $c$ and $V_{200}$ (for
the NFW halo), using a least squares fitting routine.  

\section{Modelling Results}

\begin{figure*}
\epsfxsize=0.5\hsize
\epsfbox{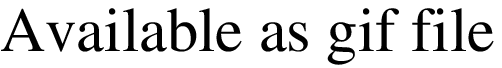}
\caption{Mass models. First and third
column show fits using the NFW halo model; second and fourth column
show models using the pseudo-isothermal halo model.  Reduced $\chi^2$
and $p$-values are given in the bottom-right of each panel.\label{decomps}}
\end{figure*}

The results of the modelling are presented in Fig.~\ref{decomps} and
Tables \ref{nfwtable} and \ref{isotable}. Fig.~\ref{decomps} displays
the models for the four $\MLstar$ assumptions, or, if no photometry
was available, only the minimum disk model. The NFW and
pseudo-isothermal (``ISO'') halo fits are shown side by side.  The
values of the halo parameters are given in Table \ref{nfwtable} for
the NFW model and Table
\ref{isotable} for the pseudo-isothermal halo model. Both Figure and Tables
also list the reduced $\chi^2$ goodness-of-fit values.  We have also
done a simple $\chi^2$ test to compute the chance $p$ that the data
and the models could originate from the same distribution. We regard
$p>0.95$ as a good fit, indicating that the data and the model
match. Values of $p<0.05$ generally indicate that the wrong model is
used to describe the data. Derived values for $p$ are shown in
Fig.~\ref{decomps}.

For the NFW fits the fitting routine often preferred very small or
negative $c$ values. These are obviously unphysical, and we have
restricted the parameter space for $c$\, to values $\ge 0.1$. When the
fit preferred a value less than 0.1 we set $c$ simply equal to 0.1,
and found the best fitting value for $V_{200}$. This is indicated in
Table \ref{nfwtable} by giving the values in italics.

\begin{table*}
\scriptsize
\begin{minipage}{120mm}
\caption{Parameters NFW models}
\label{nfwtable}
\begin{center}
\begin{tabular}{@{}lrrrrrrrrrrrrrr}
\hline
Name & $c$ & $\Delta c$ & $V_{200}$ & $\Delta V$ & $\chi^2$&$p$ && $c$ & $\Delta c$ & $V_{200}$ & $\Delta V$ & $\chi^2$ &$p$& $\Upsilon_*^{max}$\\
\hline
\noalign{\vskip 2pt}
 & \multicolumn{6}{c}{Minimum disk} && \multicolumn{6}{c}{Minimum disk + gas}\\
\cline{2-7} \cline{9-14} \\
F5631&  7.8     & 1.4& 106.8       &  10.3& 0.343 & 0.915&&  8.1     & 1.5&  99.7       &  9.4& 0.358&0.906& \\
U5750&  1.9     & 2.1& 145.7       & 122.9& 3.288 & 0.000&&  2.1     & 2.1& 122.6       & 92.0& 3.129&0.001& \\
U5005&  3.3     & 0.6& 124.6       &  13.5& 0.191 & 0.991&&  3.5     & 0.6& 107.9       & 10.2& 0.191&0.995& \\
U1230& 12.1     & 2.1&  86.4       &   6.2& 1.143 & 0.328&& 13.0     & 2.6&  79.5       &  6.5& 1.244&0.262& \\
U731 & 18.6     & 1.9&  53.1       &   2.8& 0.684 & 0.740&& 20.2     & 2.2&  48.9       &  2.6& 0.760&0.668& \\
U4173&\emph{0.1}& -  &\emph{319.6} &   -  & 0.058 & 1.000&&  2.0     & 0.7&  74.4       & 17.4& 0.116&0.999& \\
U4325&\emph{0.1}& -  &\emph{3331.6}&   -  & 1.330 & 0.132&&\emph{0.1}& -  &\emph{3203.9}&  -  & 1.132&0.257& \\
N1560&  5.3     & 0.5& 106.1       &   8.8& 2.969 & 0.000&&  5.4     & 0.5&  92.9       &  7.3& 2.630&0.000& \\
U3371&\emph{0.1}& -  &\emph{875.6} &   -  & 0.287 & 0.996&&\emph{0.1}& -  &\emph{814.7} &  -  & 0.273&0.997& \\
U628 & 12.9     & 1.2& 101.2       &   5.7& 0.347 & 0.912&& 12.9     & 1.2& 101.2       &  5.7& 0.346&0.912& \\
DDO185&\emph{0.1}& - &\emph{784.5} &   -  & 2.677 & 0.005&&\emph{0.1}& -  &\emph{651.1} &  -  & 2.326&0.012& \\
DDO189& 9.9     & 1.3& 59.2408     &   4.3& 0.171 & 0.997&& 11.0     &1.2 &  50.6    &2.8  & 0.119   &0.999&\\ 
DDO47&\emph{0.1}& -  &\emph{1332.5}&   -  & 0.320 & 0.897&&\emph{0.1}& -  &\emph{1248.2}&  -  & 0.250&0.941& \\
DDO64&\emph{0.1}& -  &\emph{1182.3}&   -  & 0.838 & 0.539&& -        & -  &   -         &  -  & -    &-    & \\
N2366& 11.2     & 2.4& 50.8487     &   7.5& 2.154 & 0.001&& 12.8     & 2.8& 41.1        & 5.8 & 2.240&0.000& \\
N3274& 30.4     & 2.1& 49.6296     &   1.6& 0.894 & 0.600&&  31.0    & 2.2& 47.7        & 1.5 & 0.830&0.685&\\ 
N4395& 12.1     & 0.9& 69.7218     &   3.8& 0.762 & 0.898&&  13.2    & 1.0& 63.3        & 3.2 & 0.733&0.927&\\ 
N4455& 5.2      & 1.0&  99.7369    &  19.3& 0.452 & 0.976&&  5.4     & 1.0& 86.8        & 14.7& 0.393&0.990&\\ 
K124 &  0.2     & 6.0& 661.1       &4320.4& 1.668 & 0.034&&   - & -  & -    &  -  & -    &-& \\
N100 &  8.3     & 1.2& 104.7       &  13.2& 0.939 & 0.018&&   - & -  & -    &  -  & -    &-& \\
U1281&\emph{0.1}& -  & \emph{785.0}&   -  & 1.376 & 0.080&&   - & -  & -    &  -  & -    &-& \\
U3137& 10.0     & 0.8& 88.2        &   2.6& 2.188 & 0.000&&   - & -  & -    &  -  & -    &-& \\
DDO52&\emph{0.1}& -  &\emph{991.0} &    - & 2.921 & 0.004&&   - & -  & -    &  -  & -    &-& \\
IC2233&\emph{0.1}&-  &\emph{1135.0}&    - & 1.256 & 0.145&&   - & -  & -    &  -  & -    &-& \\
N5023&  11.4    & 1.1& 84.6        &  7.3 & 1.044 & 0.398&&   - & -  & -    &  -  & -    &-& \\
U10310& 2.6     & 2.5& 181.3       &160.9 & 0.590 & 0.873&&   - & -  & -    &  -  & -    &-& \\
\hline
%Name & $c$ & $\Delta c$ & $V_{200}$ & $\Delta V$ & $\chi^2$ && $c$ & $\Delta c$ & $V_{200}$ & $\Delta V$ & $\chi^2$ & $\MLstar^{max}$\\
%\hline
 & \multicolumn{6}{c}{Constant $\MLstar$} && \multicolumn{7}{c}{Maximum disk}\\
\cline{2-7} \cline{9-15} \\
F5631&  6.8     & 1.6& 103.5       &  12.9& 0.360&0.904&& 1.6       & 2.8& 220.3       & 282.0& 0.373&0.895&  5.9\\
U5750&  0.8     & 3.6& 187.7       & 474.3& 3.110&0.000&& 0.6       & 4.3& 215.5       & 772.3& 3.118&0.000&  1.6\\
U5005&  2.3     & 0.6& 127.2       &  19.7& 0.170&0.997&& \emph{0.1}& -  & \emph{307.1}&   -  & 0.557&0.475&  5.4\\
U1230& 11.7     & 2.4&  77.7       &   6.6& 1.091&0.365&& 6.6       & 2.0& 73.0        &   8.7& 0.724&0.687&  6.1\\
U731 & 19.5     & 2.3&  47.6       &   2.8& 0.728&0.699&& 9.3       & 4.0& 31.8        &   6.6& 0.562&0.847& 13.5\\
U4173&\emph{0.1}& -  &\emph{176.2} &   -  & 0.129&1.000&& \emph{0.1}& -  & \emph{128.9}&   -  & 0.236&0.993&  2.4\\
U4325&\emph{0.1}& -  &\emph{2655.1}&   -  & 1.157&0.237&& \emph{0.1}& -  &\emph{1724.1}&   -  & 1.225&0.172&  4.6\\
N1560&  1.8     & 1.0& 218.9       & 105.2& 2.243&0.000&& \emph{0.1}& -  & \emph{520.6}&   -  & 8.859&0.000&  4.9\\
U3371&\emph{0.1}& -  &\emph{735.3} &   -  & 0.278&0.997&& \emph{0.1}& -  & \emph{533.1}&   -  & 0.391&0.980&  5.0\\
U628 & 12.0     & 1.3& 97.2        &   6.3& 0.341&0.915&& \emph{0.1}& -  & \emph{273.5}&   -  & 0.619&0.689&  9.8\\
DDO185&\emph{0.1}& - &\emph{575.0} &   -  & 2.269&0.014&& \emph{0.1}& -  & \emph{382.9}&   -  & 2.137&0.021&  5.0\\
DDO189& 9.6     &1.1 & 52.3        &  3.3 & 0.099&1.000&& 1.5       & 3.0& 148.6       & 220.0& 0.156&0.998&  8.3\\
DDO47&\emph{0.1}&  - &\emph{1178.0}&   -  & 0.307&0.905&& \emph{0.1}& -  &\emph{1029.8}&   -  & 0.455&0.785&  4.4\\
DDO64&\emph{0.1}&  - &\emph{966.0} &   -  & 0.770&0.615&& \emph{0.1}& -  &\emph{562.1} &   -  & 0.739&0.651&  4.1\\
N2366& 11.9     & 3.0& 36.4        & 5.8  & 1.932&0.003&& 8.6       & 7.3& 14.9        & 7.0  & 1.245&0.181&  5.4\\   
N3274& 23.7     & 2.1& 50.4        & 2.2  & 0.928&0.554&& 12.6      & 2.6& 61.8        & 7.7  & 1.804&0.013&  3.9\\
N4395& 12.6     & 1.1& 58.0        & 3.1  & 0.646&0.979&& 11.9      & 1.2& 50.8        & 3.1  & 0.577&0.994&  3.0\\  
N4455&\emph{0.1}& -  &\emph{708.3} & 32.1 & 0.630&0.849&& \emph{0.1}& -  & \emph{602.4}&  -   & 1.495&0.056&  2.3\\
\hline
\end{tabular}
\end{center}
\end{minipage}
\end{table*}

\begin{table*}
\scriptsize
\begin{minipage}{120mm}
\caption{Parameters pseudo-isothermal models}
\label{isotable}
\begin{center}
\begin{tabular}{@{}lrrrrrrrrrrrrrr}
\hline
Name & $R_C$ & $\Delta R$ & $\rho_0$ & $\Delta \rho$ & $\chi^2$ &$p$&& $R_C$ & $\Delta R$ & $\rho_0$ & $\Delta \rho$ & $\chi^2$ &$p$& $\MLstar^{max}$\\
\hline
 & \multicolumn{6}{c}{Minimum disk} && \multicolumn{6}{c}{Minimum disk + gas}\\
\cline{2-7} \cline{9-14} \\
F5631& 2.0& 0.2&  70.4& 13.1& 0.203&0.976&& 1.9& 0.2&  73.3& 14.9& 0.217&0.971&\\
U5750& 5.0& 0.9&   7.9&  1.6& 1.026&0.321&& 4.8& 1.0&   7.4&  1.8& 1.148&0.260&\\
U5005& 4.7& 0.2&  11.5&  0.8& 0.062&1.000&& 4.2& 0.2&  11.7&  0.9& 0.047&1.000&\\
U1230& 1.5& 0.3& 103.5& 38.6& 0.774&0.640&& 1.4& 0.4& 114.5& 55.2& 1.030&0.413&\\
U731 & 0.6& 0.1& 333.9& 46.0& 0.313&0.976&& 0.5& 0.1& 381.8& 69.9& 0.439&0.928&\\
U4173& 3.5& 0.2&   7.3&  0.7& 0.066&1.000&& 2.6& 0.2&   7.9&  0.8& 0.041&1.000&\\
U4325& 2.7& 0.1& 100.1&  2.1& 0.017&1.000&& 2.5& 0.1& 100.6&  2.0& 0.015&1.000&\\
N1560& 1.6& 0.1&  57.1&  4.3& 4.343&0.000&& 1.5& 0.1&  54.6&  4.8& 4.306&0.000&\\
U3371& 3.7& 0.1&  18.0&  0.3& 0.003&1.000&& 3.3& 0.1&  19.2&  0.6& 0.009&1.000&\\
U628 & 1.5& 0.2& 151.9& 27.8& 0.392&0.883&& 1.5& 0.2& 151.9& 27.8& 0.392&0.883&\\
DDO185&$\infty$&-&22.4&  2.0& 0.669&0.672&&$\infty$&-& 20.7&  2.5& 1.020&0.372&\\
DDO189&1.0& 0.1&  97.9& 11.3& 0.064&1.000&& 0.8& 0.1& 115.0& 13.7& 0.053&1.000&\\ 
DDO47& 2.1& 0.5&  47.5& 10.2& 0.205&0.975&& 1.9& 0.4&  48.4& 10.4& 0.173&0.984&\\
DDO64& 1.2& 0.2&  72.7& 11.9& 0.423&0.955&& 1.2& 0.2&  72.7& 11.9& 0.423&0.955&\\
N2366& 0.7& 0.1& 147.3& 26.4& 1.043&0.402&& 0.6& 0.1& 154.7& 37.2& 1.316&0.129&\\
N3274& 0.3& 0.1&1259.4&255.6& 0.918&0.565&& 0.3& 0.1&1340.2&290.2& 0.906&0.582&\\
N4395& 0.9& 0.1& 175.6& 18.9& 0.601&0.991&& 0.8& 0.1& 202.0& 21.8& 0.513&0.999&\\
N4455& 1.3& 0.1&  63.2&  5.1& 0.299&0.998&& 1.2& 0.1&  61.0&  5.4& 0.283&0.999&\\
K124 & 4.4& 0.2&  13.5&  0.8& 0.236&1.000&& -  & -  & -    & -   & -&-\\
N100 & 1.4& 0.1& 105.3&  4.1& 0.106&0.688&& -  & -  & -    & -   & -&- \\
U1281& 2.2& 0.1&  28.0&  1.7& 0.169&1.000&& -  & -  & -    & -   & -&- \\
U3137& 2.0& 0.2&  61.2&  8.7& 1.491&0.278&& -  & -  & -    & -   & -&- \\
DDO52& 0.7& 0.3& 123.8& 69.0& 2.193&0.052&& -  & -  & -    & -   & -&- \\ 
IC2233&3.7& 0.5&  23.0&  2.9& 1.532&0.049&& -  & -  & -    & -   & -&- \\ 
N5023& 0.9& 0.1& 200.7& 10.7& 0.277&1.000&& -  & -  & -    & -   & -&- \\ 
U10310&2.5& 0.2&  27.2&  2.2& 0.081&1.000&& -  & -  & -    & -   & -&- \\ 
\hline
%Name & $R_C$ & $\Delta R$ & $\rho_0$ & $\Delta \rho$ & $\chi^2$ && $R_C$ & $\Delta R$ & $\rho_0$ & $\Delta \rho$ & $\chi^2$ & $\MLstar^{max}$\\
%\hline
 & \multicolumn{5}{c}{Constant $\MLstar$} && \multicolumn{6}{c}{Maximum disk}\\
\cline{2-7} \cline{9-15} \\
F5631& 2.0& 0.3&  57.4& 14.3& 0.261&0.955&&  3.9& 1.4&  15.3&  8.4& 0.427&0.861& 5.9\\
U5750& 5.9& 1.7&   4.7&  1.4& 1.311&0.144&&  6.1& 1.8&   4.5&  1.4& 1.349&0.128& 1.6\\
U5005& 5.4& 0.3&   7.4&  0.5& 0.041&1.000&& 16.0& 4.2&   1.5&  0.3& 0.150&0.998& 5.4\\
U1230& 1.5& 0.4&  90.2& 43.1& 0.825&0.593&&  2.2& 0.7&  27.8& 14.9& 0.452&0.907& 6.1\\
U731 & 0.5& 0.1& 346.4& 68.0& 0.433&0.931&&  0.7& 0.3&  63.4& 42.7& 0.476&0.906& 13.5\\
U4173& 4.2& 0.6&   3.2&  0.5& 0.062&1.000&&  7.0& 2.2&   1.4&  0.4& 0.118&1.000& 2.4\\
U4325& 3.2& 0.1&  72.2&  1.1& 0.006&1.000&&  9.1& 4.7&  35.2&  2.4& 0.076&1.000& 3.8\\
N1560& 2.1& 0.2&  29.4&  2.9& 4.298&0.000&& 13.2& 4.2&   4.8&  0.4& 2.445&0.000& 4.9\\
U3371& 3.7& 0.1&  14.7&  0.7& 0.016&1.000&&  5.9& 0.8&   6.5&  0.8& 0.080&1.000& 5.0\\
U628 & 1.5& 0.2& 132.0& 29.3& 0.447&0.847&&  0.4& 0.8& 223.6&893.1& 0.946&0.460& 9.8\\
DDO185&$\infty$&-&16.9&  2.4& 1.043&0.392&&$\infty$&-&  11.6& 2.7 & 1.321&0.257& 5.0\\
DDO189&0.9& 0.1&  88.6& 10.3& 0.042&1.000&&  2.8& 0.4& 11.9 & 2.4 & 0.047&1.000& 8.3\\
DDO47& 2.4& 0.7&  37.6&  7.8& 0.152&0.989&&  7.7& 9.1&  21.3& 3.5 & 0.101&0.996& 4.4\\
DDO64& 1.4& 0.3&  52.7& 10.1& 0.413&0.959&&  2.7& 1.9&  22.0& 6.5 & 0.439&0.948& 4.1\\
N2366& 0.6& 0.1& 126.0& 37.5& 1.258&0.171&&  0.5& 0.4& 35.5 & 46.5& 1.165&0.256& 5.4\\
N3274& 0.4& 0.1& 585.3&120.7& 0.787&0.738&&  1.0& 0.2& 132.0& 33.2& 1.064&0.376& 4.0\\
N4395& 0.8& 0.1& 184.2& 23.2& 0.478&1.000&&  0.7& 0.1& 167.1& 27.0& 0.454&1.000& 3.0\\
N4455& 2.4& 0.2&  22.6&  2.0& 0.225&1.000&&  4.0& 0.7& 12.4 & 1.7 & 0.400&0.988& 2.3\\   
\hline
\end{tabular}
\end{center}
\end{minipage}
\end{table*}
 
\subsection{Comments on the mass modelling of individual galaxies}

{\bf F563-1:} The pseudo-isothermal halo fits are comparable in quality to
the NFW fits. We can compare our results with the independent fits by
dBMR.  For the minimum disk pseudo-isothermal halo we find $R_C = 2.0
\pm 0.2$ and $\rho_0 = 70.4 \pm 13.1$.
dBMR find $R_C = 1.7 \pm 0.2$ and $\rho_0 = 91.9 \pm 21.6$. These
values thus agree within their $1\sigma$ errors.  For the NFW models
our values of $c=7.8 \pm 1.4$ and $V_{200} = 106.8 \pm 10.3$ compare
well with the dBMR values of $c=10.7 \pm 1.2$ and $V_{200} = 93.1 \pm
4.3$. The $c$ values agree at the 1.1$\sigma$ level, whereas the
$V_{200}$ values agree within the $1\sigma$ errorbars.  Two completely
independent analyses of different observations on different telescopes
are thus able to reproduce the same parameters, showing that systematic effects
due to telescope pointing, smoothing etc.\ are small.

{\bf U628:} The pseudo-isothermal and NFW fits are comparable. One
should note that for this galaxy no \HI data are available, and that
the ``minimum disk'' and ``minimum disk+gas'' fits are identical. Also
no gas rotation curve is included in the ``constant $\MLstar$'' and
``maximum disk'' cases. The gas rotation curve does however usually
have only a small impact on the fit parameters, and the values
presented here will be close to the true ones.

{\bf U731:} The pseudo-isothermal models and NFW fits are comparable
except that the latter overestimate the rotation velocity in the inner
parts. Both models have trouble fitting the sharp kink at $r\sim 0.8$
kpc. This kink is artificial and caused by a lack of data between 0.8
and 4.2 kpc due to the presence of a strong night-sky line at these
velocities.

{\bf U1230:} For both models maximum disk is the best fit. This comes
at the cost, however, of a high $\MLstar(R) = 6.1$ which is hard to
explain in term of reasonable population synthesis $\MLstar$ values,
which usually find values between $\sim 0.5$ and $\sim 2$ for LSB
galaxies \citep{Hoek}.

{\bf U4173:} This rotation curve is virtually identical to the \HI
curve as the H$\alpha$ data do not add much extra constraint. Both
models fit equally well. The \HI surface density profile given in
\citet{Sw1999} gives rise to large fluctuations in the inner gas
rotation curve and was therefore smoothed slightly as explained above.

{\bf U4325:} Good fits for the pseudo-isothermal models, bad ones for
the NFW models.

{\bf U5005:} Both sets of fits are comparable. The data are however
not of high enough quality to constrain either model.

{\bf U5750:} This galaxy can again be compared with independent
observations by dBMR. We first compare the fitting parameters. For the
isothermal model we find $R_C = 5.0 \pm 0.9$ and $\rho_0 = 7.9 \pm
1.6$. dBMR find $R_C = 4.3 \pm 0.4$ and $\rho_0 = 10.6 \pm 1.0$ (again
minimum disk). These agree within the $1\sigma$ level. For the NFW
model we find $c = 1.9 \pm 2.1$ and $V_{200} = 145.7 \pm 122.9$,
compared to their $c=2.6 \pm 1.5$ and $V_{200}= 123.1 \pm 58.8$, again
in good agreement.  The large uncertainties in the NFW parameters
indicate that the rotation curve of U5750 cannot be adequately
described by that model. This is mainly caused by the very linear rise
of the inner rotation curve.  For the NFW model the only constraint is
imposed by the two \HI points at the flat part of the curve. Without
these the fit would diverge to $c=0$ and $V_{200} = \infty$. Clearly,
higher-resolution \HI observations of the \emph{outer} parts are
needed further fill in the flat part of the curve and bring down the
uncertainties of the fit parameters. This would however not change the
conclusion that this galaxy is very hard to fit with the NFW
model. Again we see that independent observations are capable of
reproducing the same set of parameters at the $1\sigma$ level.

{\bf N1560:} Due to the large number of data points and the small
scale structure present in the rotation curve, the $\chi^2$ values are
high. The pseudo-isothermal and NFW models seem to perform equally
well, though there is still a slight tendency of the NFW models to
overestimate the inner parts. Whereas the problem with rotation curves
is usually that there is not enough resolution, NGC 1560 illustrates
the opposite: the non-circular motions in the H$\alpha$ data are
large.  For the pseudo-isothermal model maximum disk fits best, but
again with a value $\MLstar(B) = 4.9$ which is much higher than one
would find on the basis of reasonable star formation histories and
stellar initial mass functions.

{\bf DDO185:} This is a very linear curve that is reasonably fitted by
pseudo-isothermal models, but poorly by NFW models. The curve is
however only of average quality, and better observations are needed to
constrain the models further.

{\bf DDO189:} This curve is equally well fitted by pseudo-isothermal
and NFW models.

{\bf DDO64:} This curve shows a bump in the inner parts, that is very
well fitted by the stellar disk in the maximum disk model, perhaps
indicating that the inner part of this galaxy is dominated by
stars. Similar bumps have been found in other highly resolved
observations of dwarf galaxies \citep[e.g.][]{Blais}.
The pseudo-isothermal model is a better fit.

{\bf N2366:} This curve is a challenge for both ISO and NFW
models. The very linear rise in the inner parts rapidly changes in a
flat part at larger radii. It is hard to reproduce such a sharp change
in slope. It is possible that the \HI curve on which the outer points
are based underestimates the rotation velocity as a comparison with
the \HI position-velocity diagram perhaps suggests. Another possible
explanation would be that non-circular motions due to the bar-like
structure in the centre affect the curve. In their study of NGC 2366
\citet{Hunter} find some weak evidence for this from the \HI velocity
field, but they also note that the rotation curve does not seem to be
affected. Clearly there is room for additional high-resolution
velocity measurements.

{\bf N3274, N4395, N4455:} The NFW models fit as well as the ISO
models.  Parts of the rotation curve of N4395 are dominated by effects
of star formation

{\bf U1281, U3137, U711, N100, DDO52, N5023, IC2233 and U10310:} For
these eight galaxies only minimum disk models are presented. For U711,
N100, DDO52 and U10310 this is because no surface photometry is
available. U1281, U3137, N100, IC2233, and N5023 are edge-on galaxies
and deriving their face-on surface brightness profiles depends on
various assumptions about the properties of their disk which would
introduce additional uncertainties in the models.

%\subsection{Chance and $\chi^2$}
\subsection{Goodness of fit}

\begin{figure}
\epsfxsize=\hsize
\epsfbox{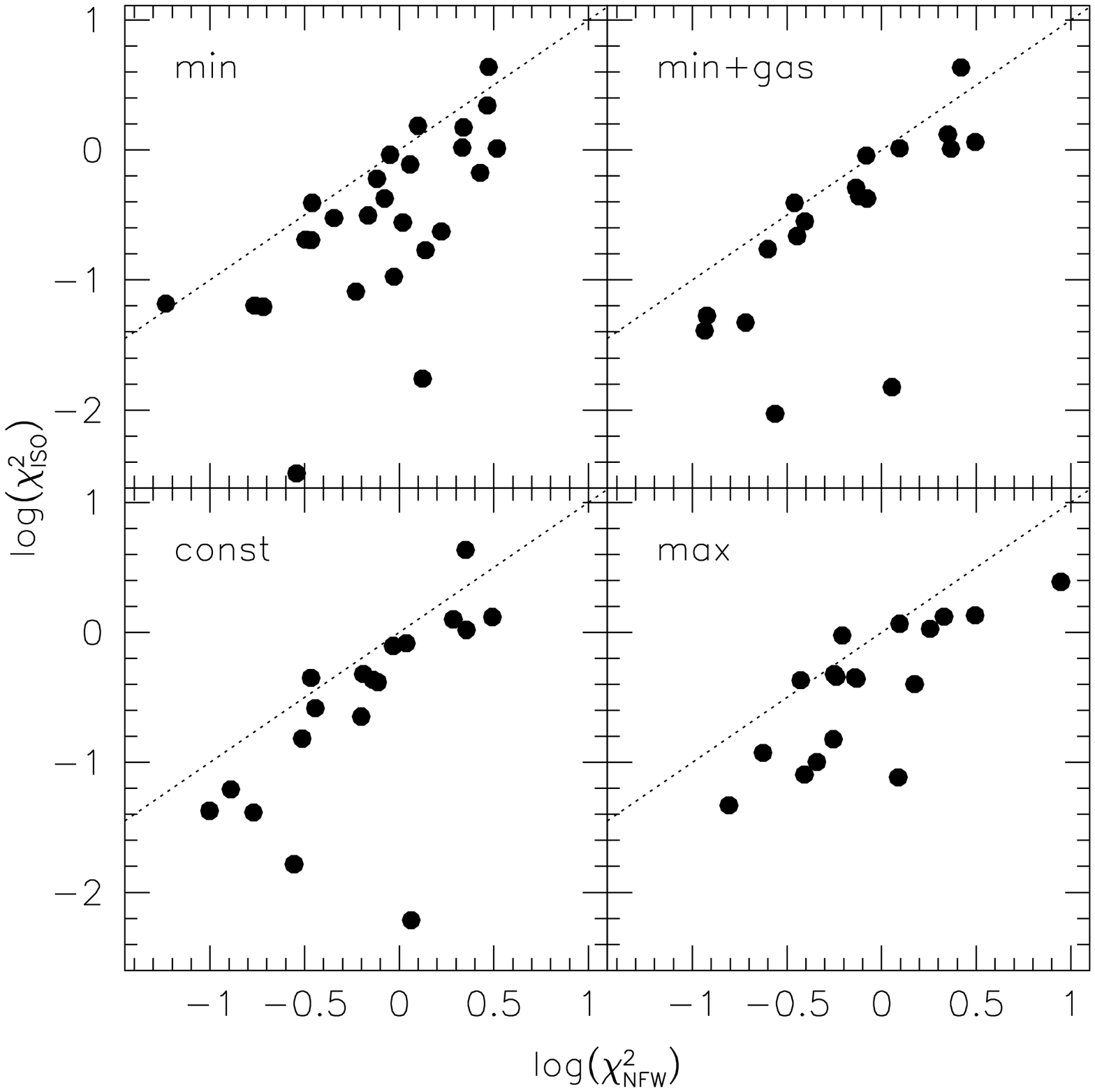}
\epsfxsize=\hsize
\epsfbox{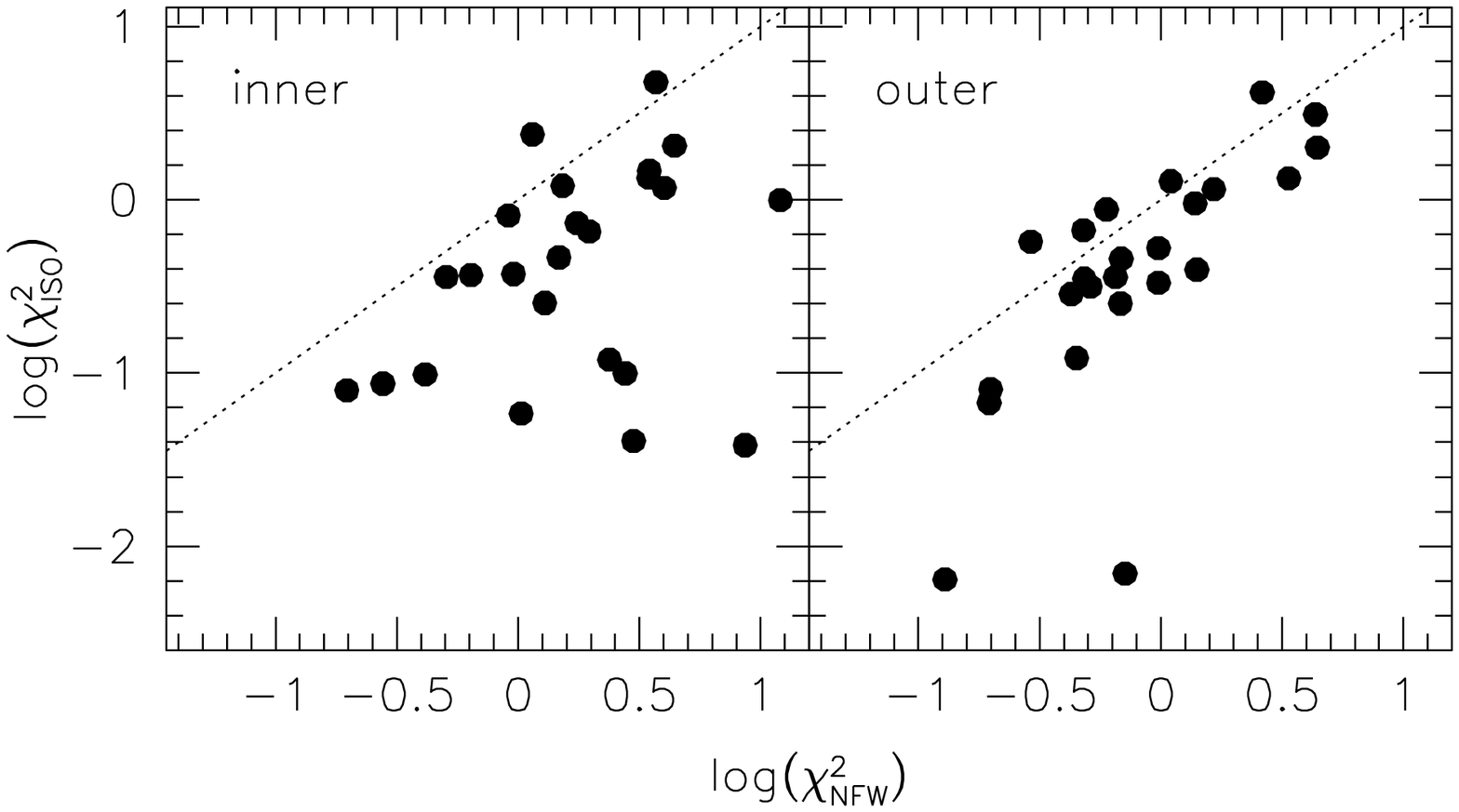}
\caption{Top: Comparison of the reduced $\chi^2$ values of mass-models using the NFW and
the pseudo-isothermal halos. In all cases the pseudo-isothermal fits
are equal to or better than the NFW fits. Dotted lines are lines of
equality. Bottom: Comparison of the reduced $\chi^2$ values for the
minimum disk case. Left panel compares the values for the inner half
of the rotation curve, right panel compares the outer half of the
rotation curve. In both cases one sees that pseudo-isothermal halos
fit better.}
\label{comparechi2}
\end{figure}

The $\chi^2$ values of the pseudo-isothermal models are generally
smaller than those of the NFW models. This is illustrated in
Fig.~\ref{comparechi2} where we plot both sets of $\chi^2$ values
against each other for the four different $\MLstar$ scenarios.  In all
cases the NFW models are either comparable or worse than the
isothermal models. We emphasise this point again in
Table~\ref{compchi2table}.  Here we list the number of galaxies with
good ($p>0.95$) and bad ($p>0.05$) fits for the two models under the
four $\MLstar$ assumptions.

We note that the $\chi^2$ values given here should not be regarded as
absolute. As described in Sect.~5, the combination of
H$\alpha$ and H{\sc i} data, the symmetrization of the curves, the
definition of the subsequent error and the imposition of a minimum
error result in a conservative estimate of the uncertainty.  The
errorbars we give here are likely an overestimate of the true
uncertainty.  This at least partly explains the very low $\chi^2$
values found for some curves.

The bottom panel in Fig.~9 compares the $\chi^2$ values for the
minimum disk case, but measured separately for the inner and outer
halves of the rotation curves, respectively. The largest discrepancies
between pseudo-isothermal and NFW are found in the inner parts, as
expected, but it is clear that also in the outer parts
pseudo-isothermal models generally provide better fits.  It is
remarkable that despite the large errorbars, there is still an obvious
preference for the pseudo-isothermal model. This conclusion is
independent of any over- or underestimate of the true error: a less conservative
estimate would even strengthen the trends observed here.

\begin{table}
\begin{center}
\caption{Comparison NFW/ISO}
\label{compchi2table}
\begin{tabular}{lllll}
\hline
        &\multicolumn{2}{c}{NFW}&\multicolumn{2}{c}{ISO}\\
        &       $p>0.95$&$p<0.05$&$p>0.95$&$p<0.05$     \\
\hline
min     &       5&      4&      11&     0\\
min+gas &       5&      5&      10&     0\\
con     &       5&      4&      10&     0\\
max     &       5&      2&      8&      0\\
\hline
\end{tabular}
\end{center}
\begin{minipage}{70mm}
NGC 1560 has been excluded from this analysis due to the
non-representative $p$-values.
\end{minipage}
\end{table}

The pseudo-isothermal model has no bad ($p<0.05$) fits, in contrast
with NFW. The absolute number of good fits is a factor 2 larger for
the pseudo-isothermal model. The proportion of ``average'' fits ($0.05
\le p \le 0.95$) is larger in the NFW case. The conclusion is that the
pseudo-isothermal halo gives a better description of the data.

\subsection{NFW halos}

\begin{figure}
\epsfxsize=\hsize
\epsfbox{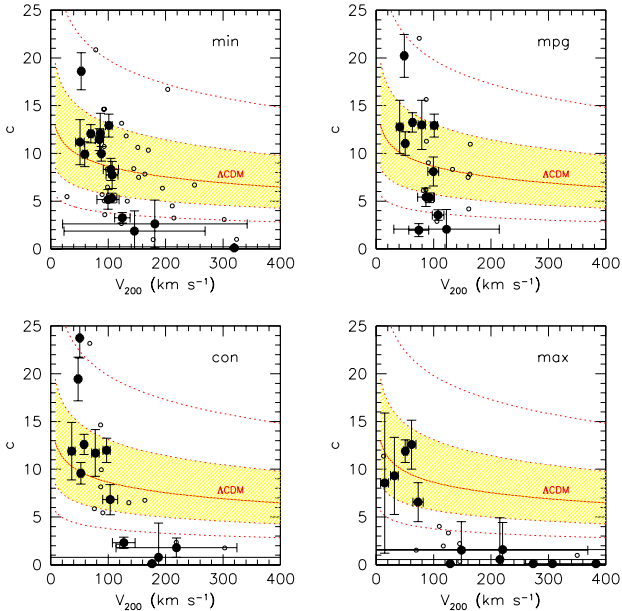}
\caption{The NFW halo concentration parameter $c$ plotted
against the halo rotation velocity $V_{200}$ for the four different
$\MLstar$ cases discussed in this paper. Large black dots are
represent our sample.  For comparison, small open dots indicate the
galaxies from the dBMR sample. The line labelled ``$\Lambda$CDM'' shows
the prediction for that cosmology derived from numerical models.  The
grey area encloses the 1$\sigma$ uncertainty (Bullock et al.\ 1999).
The upper and lower dotted line show the 2$\sigma$ uncertainty. }
\label{csumm}
\end{figure}

The mass and concentration of (numerically simulated) CDM halos depend
on the cosmological assumptions that are used as input to the
simulations.  NFW halos follow a specific relation between $c$ and
$V_{200}$ \citep{NFW-97}. Independent simulations have also given a
reasonable idea of the scatter one expects at fixed halo mass
\citep{Jing, Bullock}, even though they do not agree on the
details. In Fig.~\ref{csumm} we compare our results with these
predictions.

Shown as the full line is the prediction for a $\Lambda$ dominated
flat universe ($\Omega_m = 0.3$, $\Omega_{\Lambda} = 0.7$). There is a
general trend of increasing $c$ values towards lower $V_{200}$.  There
is some disagreement regarding the theoretical scatter in the $c$-values.
\citet{Bullock} quotes a scatter in $\log c$ of 0.18, while \citet{Jing} derives a smaller scatter in $\log c$ of 0.08. Here we adopt the larger value
from \citet{Bullock} in order to give the CDM models as much leeway
as possible.  The dotted lines in Fig.~\ref{csumm} indicate the
1$\sigma$ (grey area) and $2\sigma$ scatter where $\sigma =
\sigma(\log c) = 0.18$.
The small dots represent the values for other LSB and dwarf galaxies
derived in dBMR.  Our data show a similar behaviour, though the locus
of our data seems to be better defined than that of the dBMR
data.  The spread in $c$ larger than the models predict, which is best
illustrated by the excess of low-$c$ points.

\section{Pointing effects}

In an analysis of this nature it is crucial that the slit is aligned
with the center of the galaxy. Any off-set from center will lead to a
rotation curve that is less steep than a properly centered rotation
curve unless the entire galaxy is a pure solid body rotator.
It is therefore theoretically conceivable that systematic
effects may have caused us to underestimate the inner slope of the
rotation curve, thus mistaking the incorrectly lowered slope of the
NFW profile for that of an isothermal halo.

Here we present several arguments why this is not the case, and why the
observationally derived slopes are very close to the true slopes.

\subsection{Evidence from observations}
As already mentioned in Sect.~2, we took considerable care when
acquiring the galaxy at the telescope. Most of the nearby galaxies in
our sample were visible in the guiding-camera, and they were used to
test our off-set procedure. This involved re-acquiring a galaxy
several times in order to test the stability of the procedure, as well
as moving back and forth to the acquisition star to test for
repeatability. We found the stability of the system to be better than
$1''$. Of course, for the faintest galaxies, we rely entirely on the
position given for the center, which we took from \citet{Sw1999} or
\citet{Stil} or the NED database.

\begin{figure}
\epsfxsize=\hsize
\epsfbox{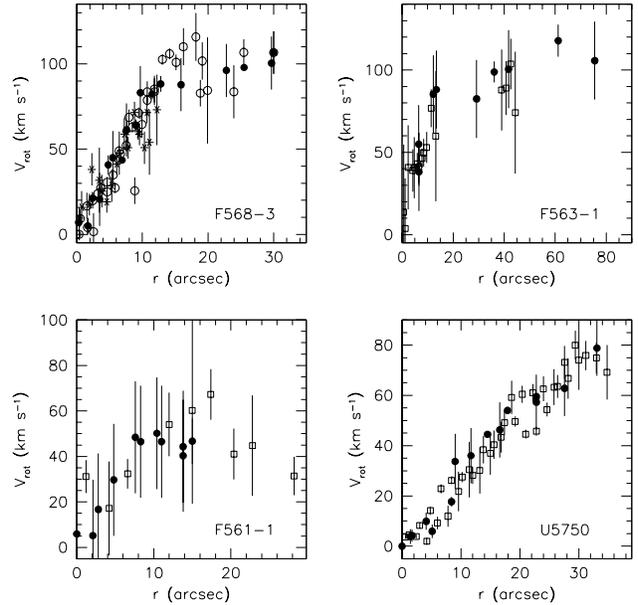}
\caption{The rotation curves derived from independent H$\alpha$ 
observations for four LSB galaxies. Open squares are the data
presented here. Solid circles are from \citet{MRdB}. Open circles in
the top-left panel (F568-3) are from \citet{Sw2000} and stars are from
\citet{Pick98}. Agreement between independent observations in
good.}\label{compcurve}
\end{figure}

There are no systematic differences between the rotation curves
acquired by us at the 1.93m telescope at OHP, by \citet{MRdB} at the
Kitt-Peak 4m telescope, and the Las Campanas du Pont 2.5 telescope, by
\citet{Sw2000} at the 200'' Hale Telescope at Mt.~Palomar and 
by \citet{Pick98} at the MMT.  This indicates that either all or none
of the data sets suffer from systematic effects due to mispointing. A
number of galaxies have been observed multiple times by different
observers, and similarly show no systematic differences.  This is
illustrated in Fig.~\ref{compcurve} where the raw data of four LSB
galaxy rotation curves are compared. In one case there are three
independent data sets available, all agreeing with each other.  It is
unlikely that independent observers miss the center of the
galaxy by the same amount in the
same direction every time.

If the true rotation curves were really bona-fide NFW curves we would
expect that with the increased resolution of the H$\alpha$ data, at
least some of them should start to look more like NFW curves. After
all, the combined samples in this paper and \citet{MRdB} contain a few
dozen rotation curves. Even in the presence of (hypothetical) pointing
effects, some should have hit the nucleus and have shown the
characteristic NFW shape. This has not happened.  The situation is
quite the reverse: the best resolved rotation curves show the flattest
cores (see Sect.~9 and \citealt{dBMBR}).

This leads to an interesting contradiction: if indeed we imagine that
a pointing offset leads to a systematic under estimate of the slope,
we ought to see a trend of rotation curve shape with distance. This is
not seen. A pointing offset of a given number of arcseconds
corresponds to a different physical scale in each galaxy, depending on
its distance.  Our sample and that of \citet{MRdB} span a range in
distance of a factor of $\sim 200$. If we imagine that a pointing
offset of, say, 0.5$''$ causes us to mistake an NFW halo for a
pseudo-isothermal one in the distant galaxies in our sample, then this
implies that in order to mislead ourselves in a similar way, we ought
to be making an error of almost 1.5$'$ in our pointing when observing
the nearest galaxies.  Though this is an extreme example, it is highly
improbably that systematic effects in target acquisition at the
telescope depend on the distance of the target.

Assuming then that we managed in the large majority of the cases to
indeed home in on the center of the light distribution of the galaxy,
one could still assume that the dynamical center of the galaxy does
not coincide with the center of the light distribution. Centering on
the light distribution would thus not give us the true rotation
curve. This has of course just the same effect of mis-pointing, but
now the fault lies with the galaxy, rather than the observer.

To test this one needs high-resolution velocity fields, to which one
can then fit a tilted ring model. The tilted ring model has the
central position of the rings as a free parameter, and will thus
immediately show whether there is a significant discrepancy between
the optical and the dynamical center or not.  This is precisely what
\citet{Blais}  have done. They have derived H$\alpha$
rotation curves for a number of late-type galaxies using full 2D
Fabry-Perot velocity fields. They find the same slowly-rising rotation
curves, and conclude that NFW models are incompatible with their data,
which prefer a core-dominated halo or a model with a shallow inner
density slope.

This is the same conclusion as reached by \citet{Bol}, who obtained
high-resolution $5''$ CO observations of the nearby dwarf galaxy NGC
4605. They derived a rotation curve from a 3D data cube, again using a
tilted ring model, and find a slowly rising rotation curve
incompatible with NFW.

A number of galaxies in our sample is edge-on, and it is conceivable
that, despite contrary evidence, these data are affected by optical
depth or projection effects. These edge-on galaxies are however only a
small fraction of the galaxies investigated. For example, of the 56
galaxies presented here and in \citet{dBMR}, only 16 have inclinations
larger than 70$^{\circ}$.  Though there are no systematic differences
between the $\sim 30$ per cent with $i\geq 70^{\circ}$ and the $\sim
70$ per cent with $i<70^{\circ}$, one could easily disregard the
high-inclination galaxies without affecting any of the
conclusions. There are no significant trends of $c$ or slope with
inclination.

In summary, there is no evidence that the data set presented in this
paper suffers from significant systematic effects, nor is it
inconsistent with any other relevant data set in the
literature. Pointing effects play no significant role, and the
observed rotation curves represent the overall
dynamics of our galaxies well.

\subsection{Evidence from models}

To quantify the effects of mispointing we have created model NFW
velocity fields and have derived rotation curves for various pointing
offsets.  We consider model velocity fields assuming NFW halos with
$i=60^{\circ}$, $V_{200} = 100$ km s$^{-1}$ and $c=4, 9, 20$,
respectively. Each of these three velocity fields is ``observed'' with
a 1$''$ wide slit, offset by $d = 0'', 2'', 5'', 10''$ from the
center, parallel to the major axis.  We assume a distance of 10 Mpc,
which is the average distance of the large majority of galaxies in our
sample (disregarding the 4 galaxies with $D>40$ Mpc.)

\begin{figure}
\epsfxsize=\hsize
\epsfbox{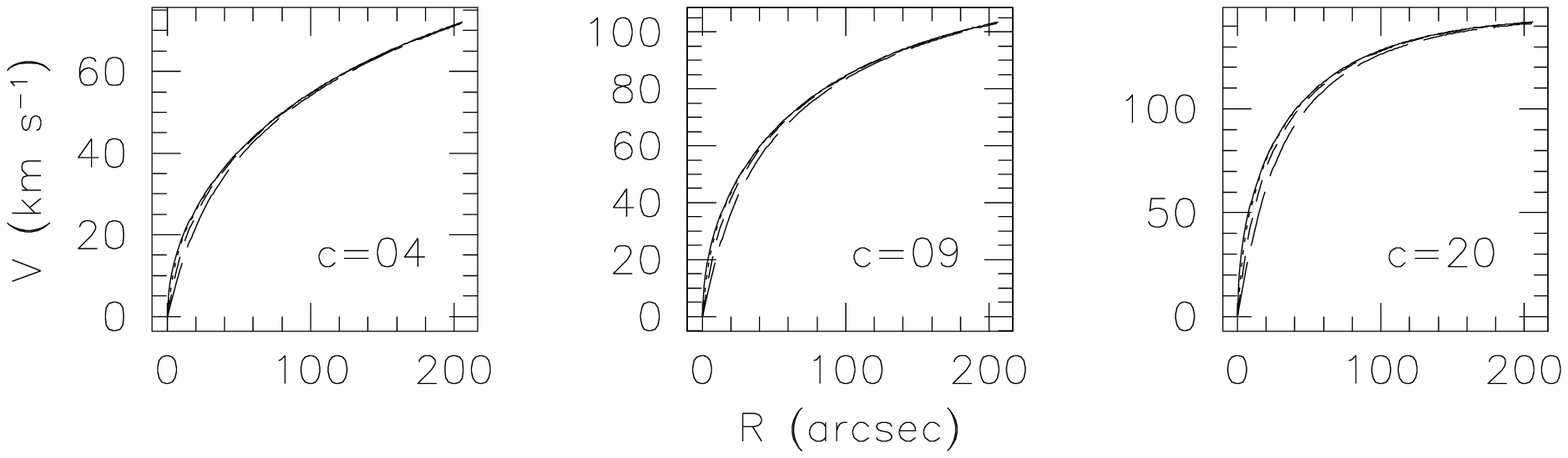}
\vspace{-5pt}
\epsfxsize=\hsize
\epsfbox{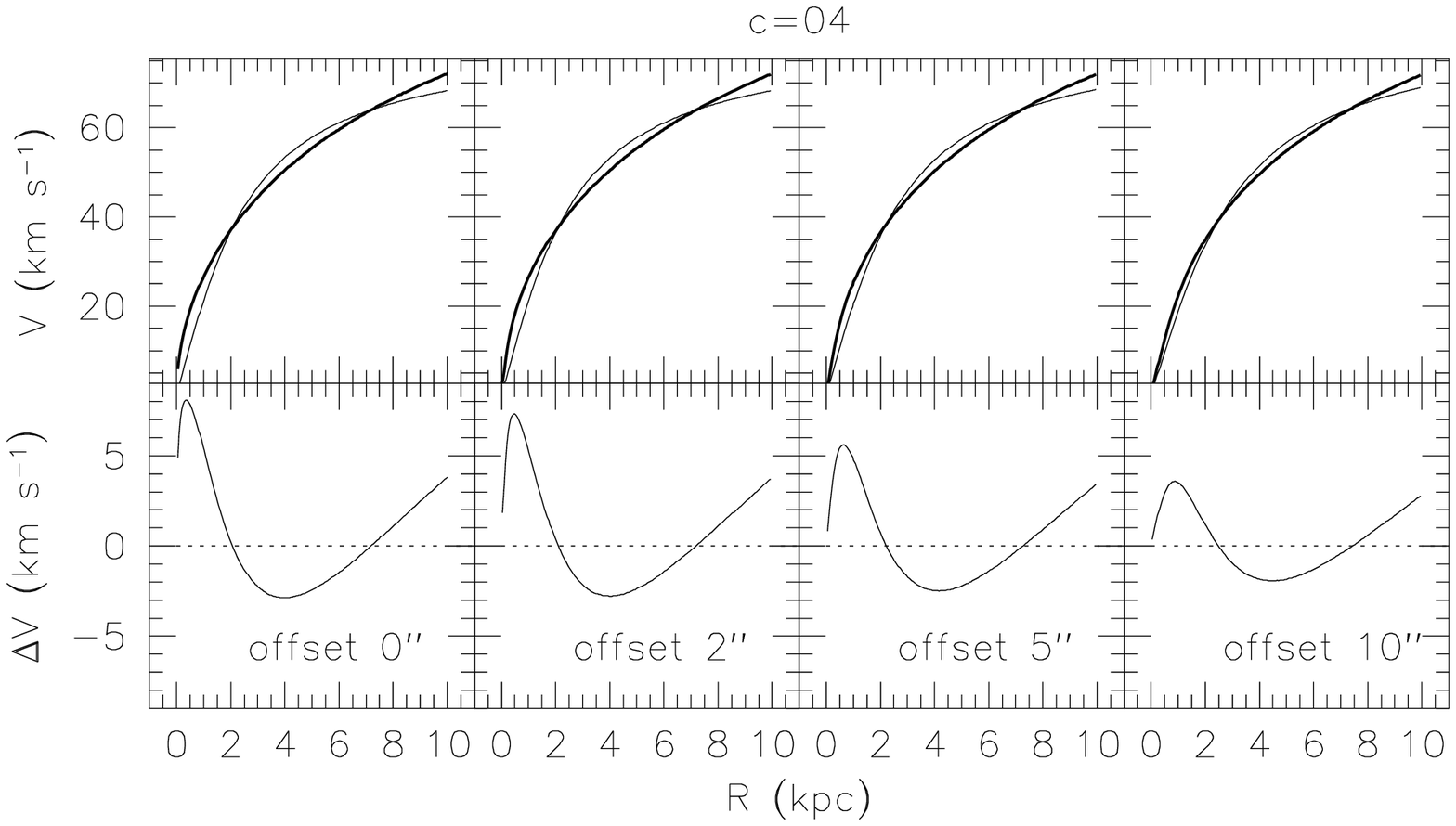}
\vspace{-5pt}
\epsfxsize=\hsize
\epsfbox{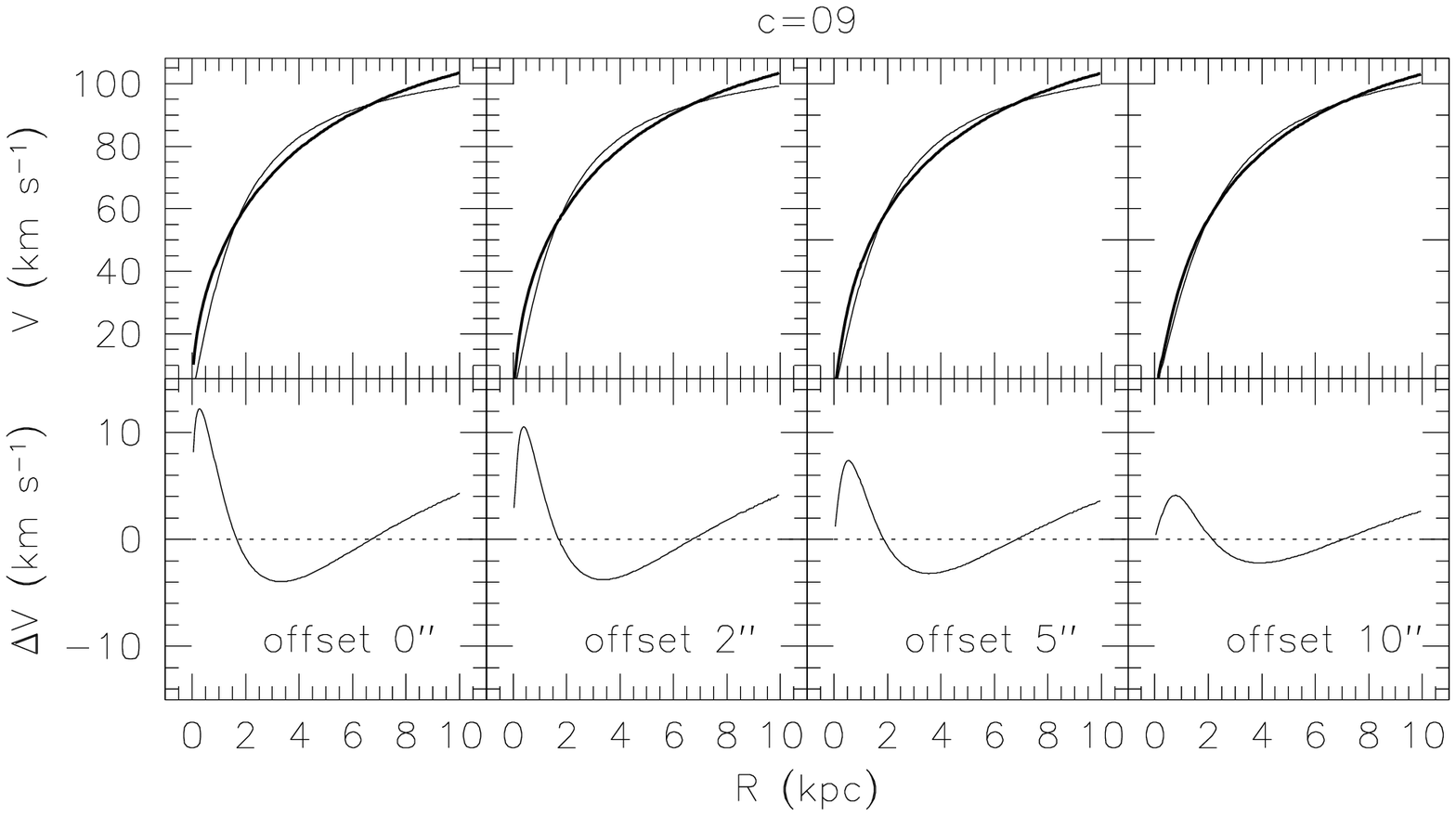}
\vspace{-5pt}
\epsfxsize=\hsize
\epsfbox{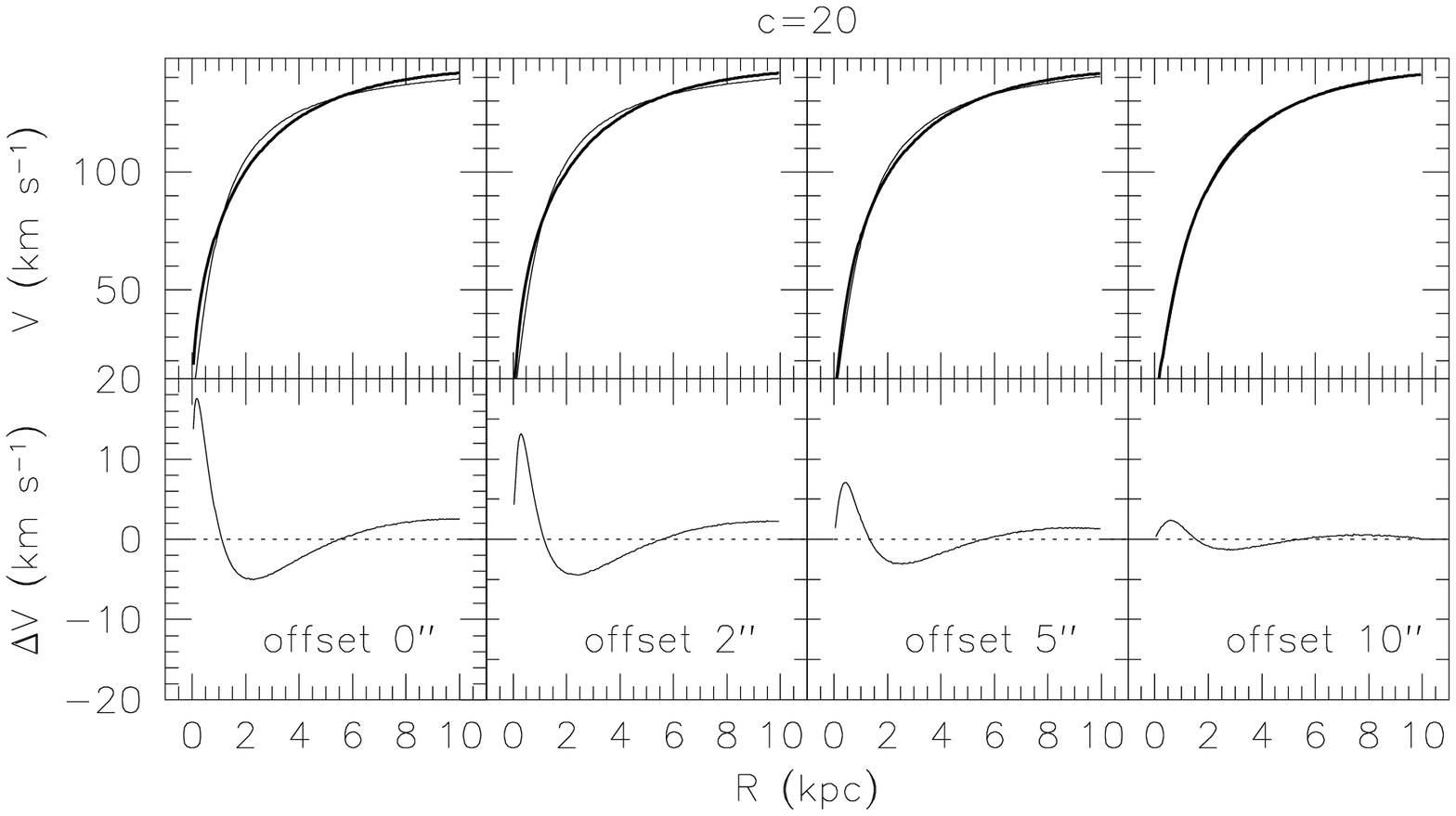}
\caption{Top panel: NFW model curves assuming $D=10$ Mpc, 
$V_{200}=100$ km s$^{-1}$, and $c=4$ (left), 9 (middle) and 20
(right). For each $c$ we derived curves for various offsets:
$0''$ (full line), $2''$ (dotted), $5''$ (short-dashed) and $10''$
(long-dashed). The other panels show the rotation curves for
each combination of $c$ and offset (thick curve). Overplotted are the
best fitting pseudo-isothermal models (thin curve). Also shown are the
differences between the two models. 
}\label{modelvel}
\end{figure}

The top panel in Fig.~\ref{modelvel} shows the rotation curves, for
each combination of $c$ and $d$. The most important question we
want to answer here is under what conditions an NFW velocity field
can produce a rotation curve that mimics a pseudo-isothermal curve.

The three bottom panels in Fig.~\ref{modelvel} show the derived offset
NFW curves, with the best fitting pseudo-isothermal models
overplotted. Also shown are the differences between both types of
curves. It is clear that in almost all cases the shape of the
pseudo-isothermal halo curve is distinctly different from that of the
NFW curve. The only cases where one might mistake an offset NFW curve
for a pseudo-isothermal curve are those for ($c=9$; $d=10''$) and
($c=20$; $d=5'', 10''$). Pointing errors of this magnitude are simply
not present in the data.  If we accept an uncertainty of at most a few
tenths of an arcsecond, it is easy to see that pointing effects are
only likely to affect galaxies at distances $\ga 100$ Mpc.

We can also regard the NFW curves in Fig.~\ref{modelvel} as best fits
to the plotted pseudo-isothermal curves. In this case we can make a
direct comparison with the fits plotted in Fig.~\ref{decomps}, and see
that the characteristic overprediction of the inner part of the curve
is also present in the data. Except for the three cases mentioned
above, Fig.~\ref{modelvel} shows that in all cases the difference
between NFW and pseudo-isothermal is most pronounced in the inner 2
kpc with residuals between $\sim 5$ and $\sim 20$ km s$^{-1}$. The
majority of galaxies in Fig.~\ref{decomps} does show such residuals,
meaning that even in the presence of (hypothetical) modest pointing
offsets, NFW curves would look different from the curves observed and
shown in Fig.~\ref{decomps}.

In summary, for the galaxies in this sample the effect of missing the
centers of the galaxies would not be strong enough to masquerade NFW
curves as pseudo-isothermal curves, except in the case of very
(unrealistically) large offsets ($\ga 5''$). This conclusion, combined
with the conclusion from Sect.~8.1 that any pointing errors must be
below the $1''$ level, shows that the observed rotation curves must be
close to the true rotation curves. Systematic observational effects
cannot hide NFW halos in LSB galaxies.

\section{Mass profiles}
\subsection{Introduction}

In \citet{dBMBR} (hereafter dBMBR) the rotation curve sample presented
in dBMR and part of the sample (the Jan00 galaxies) presented in this
work were used to calculate the mass-density profiles that give rise
to the observed rotation curves. As the different halo models predict
different shapes of the mass distribution this gives one a direct test
of the applicability of a particular model.

The mass density profile is derived by assuming a spherical mass
distribution. Then, from
%\begin{equation}
$\nabla^2 \Phi = 4\pi G \rho$
%\end{equation}
and 
%\begin{equation}
$\Phi = -{GM}/{r}$
%\end{equation}
one can derive  the mass density $\rho(r)$:
\begin{equation}
4\pi G\rho(r) = 2\frac{v}{r} \frac{\partial v}{\partial r} +
\left(\frac{v}{r}\right)^2.
\end{equation}
In the above $v$ is the rotation velocity and $r$
is the radius.

This inversion is only valid if the contribution by the gas and stars
is negligible, i.e.\ one implicitly assumes a situation (close to)
minimum disk. It is well-established that this is a reasonable
assumption for LSB (dwarf) galaxies.  A minimum disk also provides an
upper limit on the steepness of the slopes of the halo mass-density
profiles: inclusion of gas and stars will necessarily
tend to flatten the slopes.

dBMBR find that the shape of the minimum-disk halo profiles can
usually be characterised by two power-law components of the form $\rho
\sim r^{\alpha}$. The outer slope has a value close to $\alpha = -2$,
whereas the inner slope is more shallow.  The models tested here make
distinct predictions regarding the value of the inner slope: the
pseudo-isothermal halo predicts $\alpha = 0$, the NFW halo predicts
$\alpha = -1$, and more recent CDM simulations \citep{moore99} find
even steeper values of $\alpha=-1.5$.

dBMBR find that the distribution of the inner slopes $\alpha$ is
asymmetric, with a well-defined peak at $\alpha = -0.2 \pm 0.2$,
inconsistent with CDM predictions.  This distribution has a broad wing
however, extending to values of $\alpha < -1.8$, seemingly consistent
with the steep slopes demanded by CDM.  dBMBR show that the inner
slope one derives depends on the radius of the innermost sampled point
$r_{\rm in}$ of the rotation curve. Larger values of $r_{\rm in}$
sample the range of radii where the NFW profile and the
pseudo-isothermal profile have similar slopes (which does not imply
they have similar \emph{shapes}!). Small values of $r_{\rm in}$ probe
the region where the predicted slopes are distinctly different. The
galaxies with small values for $r_{\rm in}$ indeed show the clearest
evidence for the presence of a constant-density core (i.e.\
$\alpha\sim 0$).  dBMBR conclude that all data are consistent with LSB
galaxies having core-dominated halos with core-radii of a few kpc.

The implication is that galaxies with large values of $r_{\rm in}$ and
consequently with steep slopes, should show much shallower slopes once
the resolution is increased (and $r_{\rm in}$ decreased).
The current data set puts us in an excellent position to test this
prediction.  For a large number of galaxies we now have
high-resolution rotation curves, as well as independent
lower-resolution curves from \citet{Sw1999}. We should thus see a
systematic decrease in inner slope when moving from the pure \HI data
to the high-resolution data.

\subsection{Results}

\subsubsection{Minimum disk profiles}
Fig.~\ref{massprofiles} plots the derived mass profiles for our
complete sample.  It is clear that most of the galaxies from our Feb01
run, which were specifically chosen to be nearby objects, are also
characterised by an almost flat inner core with a radius of a few kpc,
in contrast with the steep $\alpha=-1.5$ power-law slope demanded by
CDM, or even the inner $\alpha=-1.0$ power-law slope of the NFW
profile.

Table~\ref{slopetable} lists the values of the slopes derived using
the method described in dBMBR. In short, after determining the
``break-radius'' where the slope changes most rapidly, we determined
the slope of the inner component using a weighted least-squares
fit. The uncertainty was determined by re-measuring the slope twice,
once by including the first data point outside the break-radius, and
once by omitting the data point at the break-radius. The maximum
difference between these two values and the original slope was adopted
as the uncertainty.  Following dBMBR we re-plot the values of the
inner slope against the value of $r_{\rm in}$, this time also including
the Feb01 data (see Fig.~\ref{sloperadius}).  The new galaxies are
consistent with a core, not a cusp, and the galaxies with small values
of $r_{\rm in}$ have flat mass density slopes.

\subsubsection{The case of N3274}

The only exception seems to be N3274 which shows a slope close to $-1$
even though its value for $r_{\rm in}$ is quite small.  The
pseudo-isothermal halo fit shows that this galaxy has the smallest
(minimum disk) core-radius and highest central density of the entire
sample. Though seemingly consistent with the NFW profile, it is also
consistent with a pseudo-isothermal halo profile with a small core
radius. Furthermore, it has by far the highest surface brightness and
smallest scale-length of the entire sample, and in this case it is
very likely that the minimum disk assumption breaks down and the
stellar component \emph{is} important in the inner parts.

\begin{table}
\begin{center}
\caption{Inner power-law slopes $\alpha$}
\label{slopetable}
\begin{tabular}{lrrrrr}
\hline
Name    & $\alpha_{min}$ & $\Delta \alpha$ &$r_{\rm in}$ (kpc)& $\alpha_{con}$ & $\Delta \alpha$\\
\hline
F563-1& --0.01& 0.70&    0.55&  0.21 &1.38 \\
U628  & --1.29& 0.08&   0.95& --1.37 &1.37 \\
U711  & --0.12& 0.07&   0.38& -- & -- \\
U731 &  --0.52& 0.45&   0.35& --0.44 &0.15 \\
U1230&  --0.13& 0.26&   0.74&  0.08 &0.47 \\
U1281&  --0.04& 0.01&   0.08& -- & -- \\
U3137&  --0.20& 0.10&   0.27& -- & -- \\
U3371&  --0.16& 0.10&   0.56& --0.02 &0.19 \\
U4173&  --0.77& 0.13&   0.73& --0.26 &0.46 \\
U4325&  --0.33& 0.03&   0.15& --0.24 &0.06 \\
U5005&  --0.58& 0.09&   0.76& --0.53 &0.24 \\
U5750&  --0.17& 0.27&   0.81&  0.26 &0.71 \\
N100 &  --0.19& 0.17&   0.19& -- & -- \\
N1560&  --0.26& 0.26&   0.04& --0.04 &0.24 \\
N2366&   0.24& 0.13 &   0.09&  0.45 &0.45 \\
N4395&  --0.40& 0.07 &   0.05& --0.52 &0.02 \\
N3274&  --0.90& 0.13 &   0.10& --0.67 &0.17 \\
N4455&  --0.57& 0.21 &   0.10& --0.70 &0.25 \\
U10310&  0.10& 0.36 &   0.22& -- & -- \\ 
N5023 & --0.39& 0.14 &   0.07& -- & -- \\ 
IC2233& --0.20& 0.22 &   0.15& -- & -- \\ 
DDO52 &  0.34& 0.50 &   0.14& -- & -- \\ 
DDO64 & --0.21& 0.11 &   0.09& --0.16 &0.58 \\
DDO47&  --0.42& 0.25 &   0.27& --0.28 &0.30 \\
DDO185& --0.18& 0.29 &   0.07& --0.23 &0.61 \\
DDO189& --0.82& 0.05 &   0.46& --0.87 &0.35\\
\hline
\end{tabular}
\end{center}
\end{table}

\subsubsection{Constant $\MLstar$ profiles}

We tested this by re-deriving the slopes for the constant $\MLstar =
1.4$ case (see dBMR for a justification of this value). The halo
rotation curve was found by quadratically subtracting the gas-rotation
curve and the rotation curve of the stars. This treatment is likely to
be too naive, as in a non-minimum disk case one expects the disk to
influence the dark matter distribution to some degree (though perhaps
not for LSB galaxies). A full treatment of this problem is beyond the
scope of this paper, and the naive procedure suffices to illustrate
the main point: as the stellar mass-to-light ratio is increased the
inner slope of the halo mass-density profile becomes flatter.

The slopes for the constant $\MLstar$ case are listed in
Table~\ref{slopetable}. In Fig.~\ref{slopecon} we again plot the
derived slopes against the inner radii for the galaxies in our sample
where a constant $\MLstar$ model was available.  A comparison with
Fig.~\ref{sloperadius} shows that the data points have all moved up,
as expected. In a number of cases we found rather large positive
slopes, which would imply that these galaxies have hollow halos. This
is rather improbable, and a more realistic explanation is that the
value $\MLstar = 1.4$ is an overestimate of the true $M/L_*$.

N3274 has a slope of $-0.66$ in the constant $\MLstar$ case, which is
consistent with the slope one would expect for a halo with a
core-radius of a few tenths of a kpc. Though this galaxy obviously
cannot be used to prove or disprove either model, it is clear that
galaxies which show steep inner slopes are likely to be of high
surface brightness with inner regions dominated by stars.  In order to
unambiguously prove that a galaxy has a bona-fide NFW halo and is
inconsistent with the pseudo-isothermal halo model, it needs 
to have a steep inner slope, \emph{and} a small value of $r_{\rm in}$
\emph{and} a low surface brightness.  Such galaxies have not been found yet.

\begin{figure*}
\epsfxsize=\hsize
\epsfbox{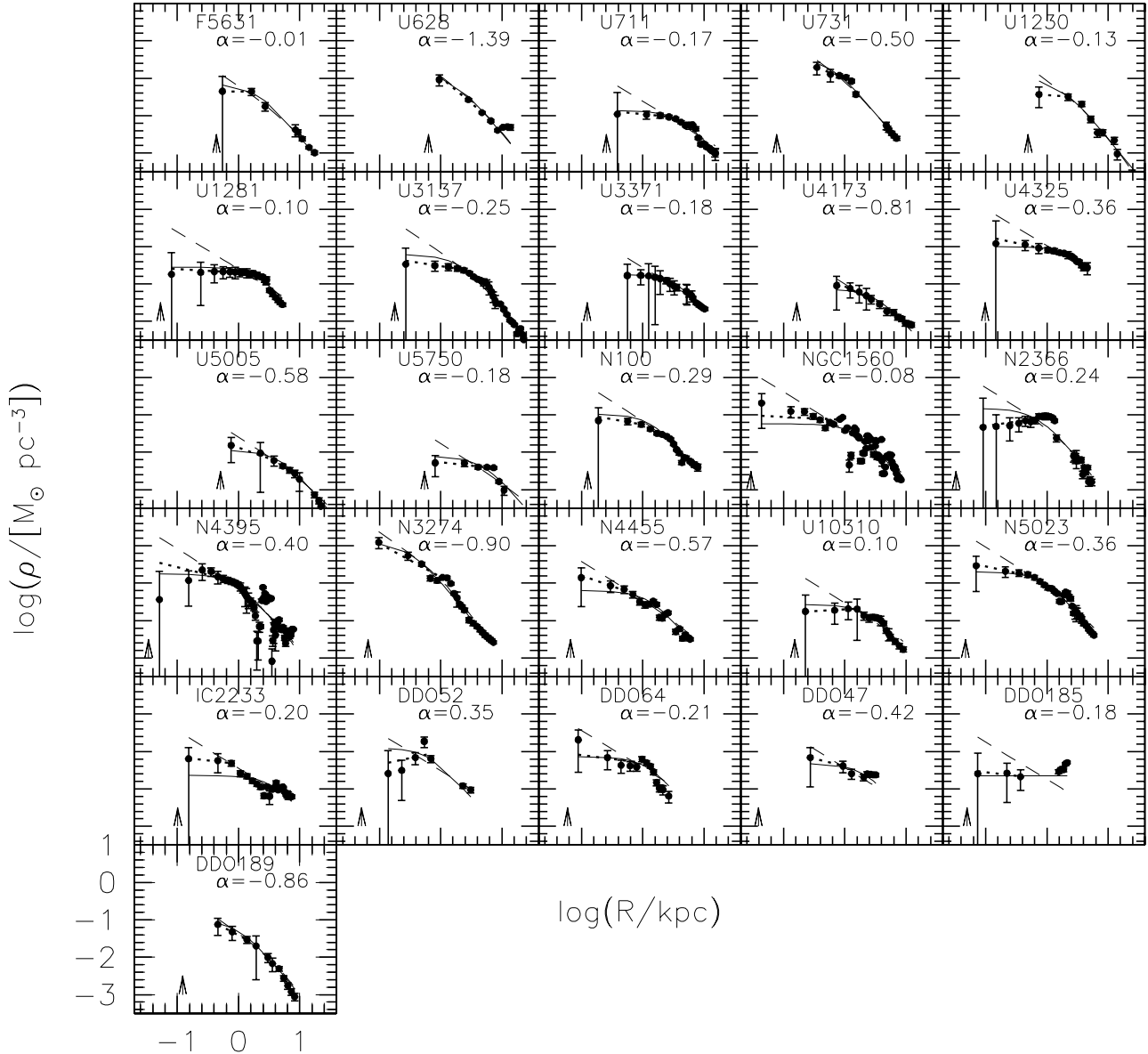}
\caption{Mass profiles of our sample galaxies derived from the 
high-resolution rotation curves. The profiles can be characterised by
a steep $r^{-2}$ outer component, and a more shallow inner component
(``core''). Also shown are the mass density profiles implied by the
best-fitting minimum disk models. Shown are the pseudo-isothermal halo
(full line) and the NFW halo (long-dashed line). We have also fitted a
power-law to the inner shallow part (thick short-dashed line). The
slope $\alpha$ is given in the top-left corners of the panels. All
panels are at the same scale (denoted in bottom-left corner). Galaxies
are labelled with their name. The arrows indicate an angular
size of 2$''$, the typical value of the seeing.}
\label{massprofiles}
\end{figure*}

\begin{figure}
\epsfxsize=\hsize
\epsfbox{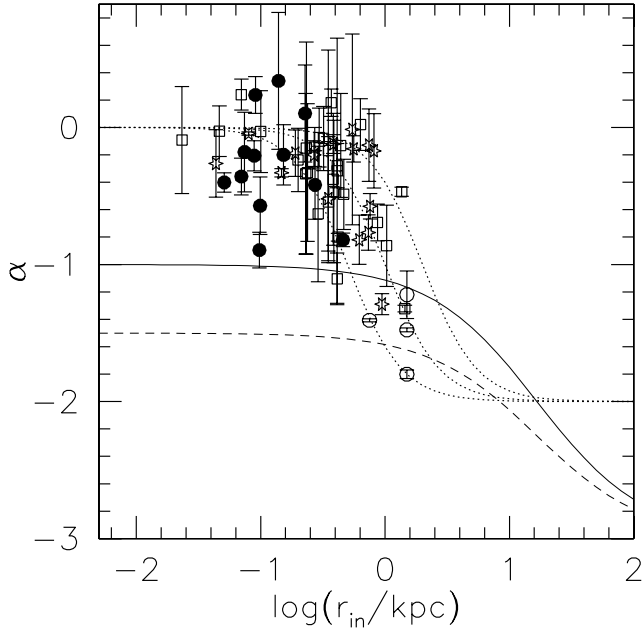}
\caption{Value of the inner slope $\alpha$ of the mass-density
profiles plotted against the radius of the innermost point. Open
circles are from the dBMR sample, stars are the Jan00 part of our
sample, as published in dBMBR.  Filled circles indicate the new
galaxies from the Feb01 part of the sample.  Over-plotted are the
theoretical slopes of a pseudo-isothermal halo model (dotted lines)
with core radii of 0.5 (left-most), 1 (centre) and 2 (right-most) kpc.
The full line represents a NFW model, the dashed line a
CDM $r^{-1.5}$ model. Both of the latter models have
parameters $c=8$ and $V_{200} = 100$ km s$^{-1}$, which were chosen to
approximately fit the data points in the lower part of the diagram.}
\label{sloperadius}
\end{figure}

\begin{figure}
\epsfxsize=\hsize
\epsfbox{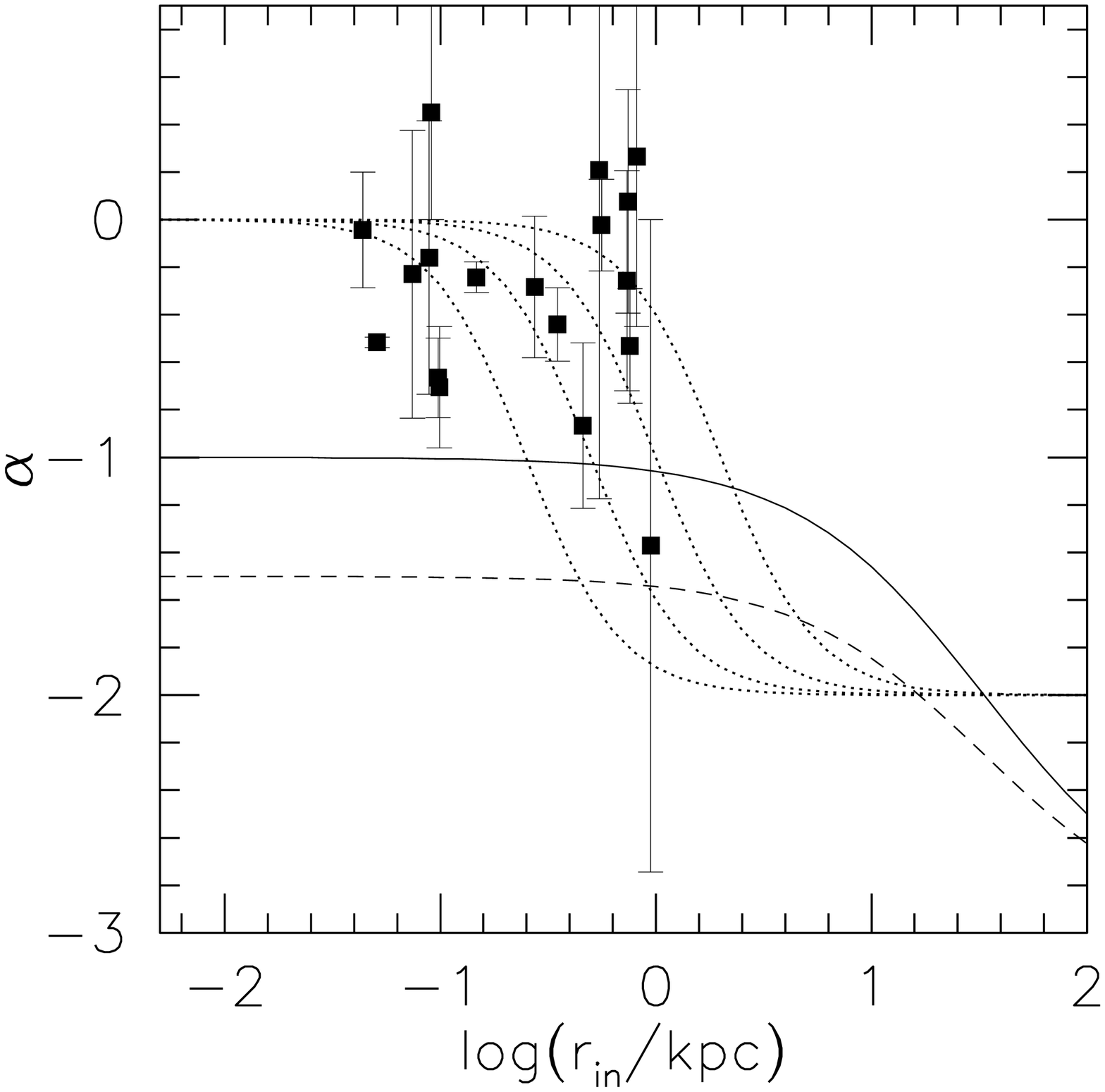}
\caption{The inner power-law slopes $\alpha$ for the constant $M/L_* = 1.4$ assumption. 
\label{slopecon}}
\end{figure}

\subsubsection{Resolution and slope}
One of the conclusions of dBMBR is that as the resolution of rotation
curves ) is increased (i.e.\ $r_{\rm in}$ is decreased), the derived
inner slope should become shallower. We test this by comparing the
slopes and inner radii for a number of curves for which we have
high-resolution curves and medium-resolution ($\sim 15''$) \HI curves.
We determined the slopes of both the high-resolution curves and the
\HI curves.

\begin{figure*}
\epsfxsize=\hsize
\epsfbox[21 160 569 510]{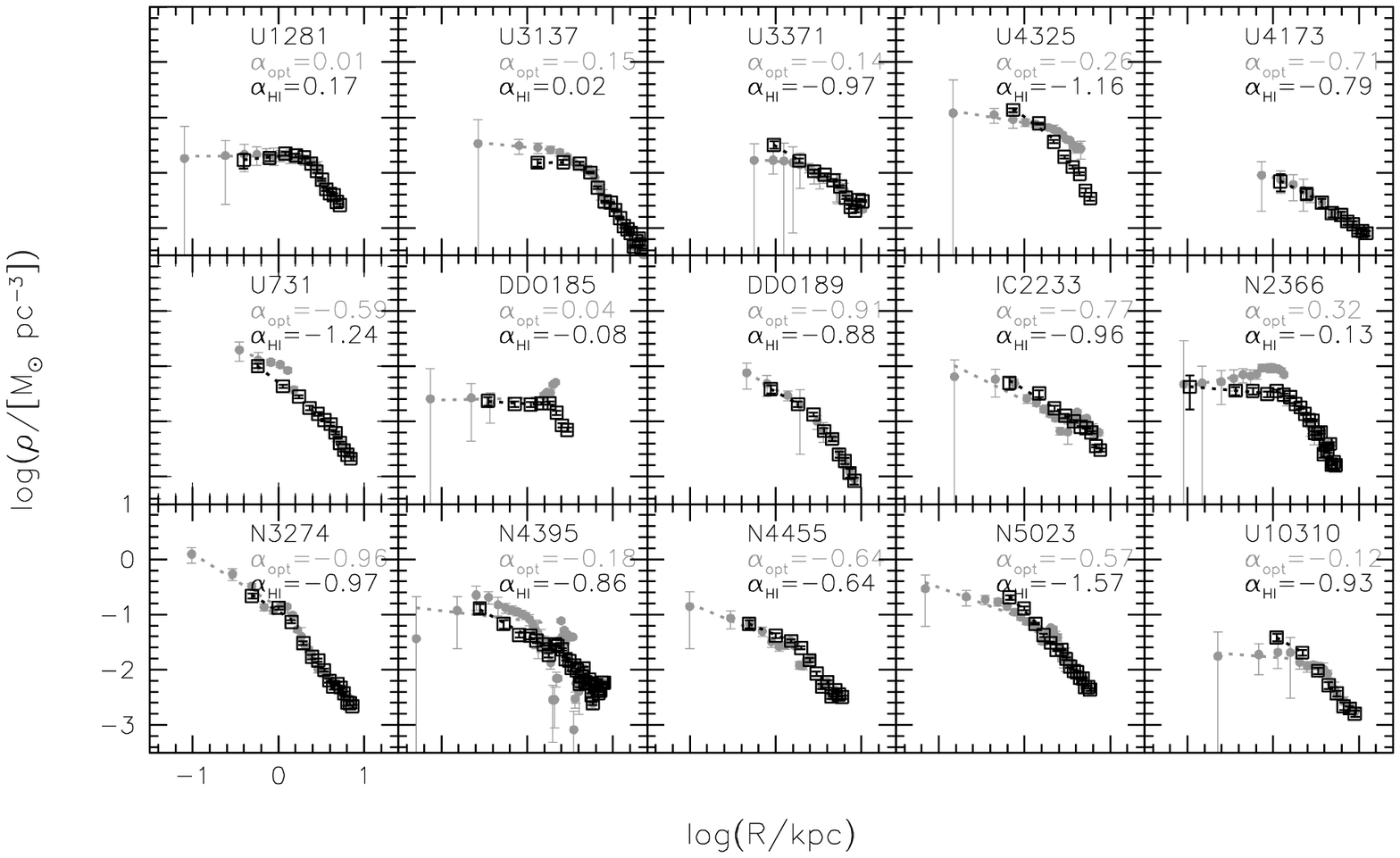}
\caption{Comparison of mass density profiles for a number of galaxies as derived 
from the high-resolution curves (grey data points) and from medium
resolution \HI curves (black points). The dashed lines indicate the
range over which the inner slope is fitted. Derived values are
indicated in each panel. }
\label{hiprofiles}
\end{figure*}

\begin{figure}
\epsfxsize=\hsize
\epsfbox{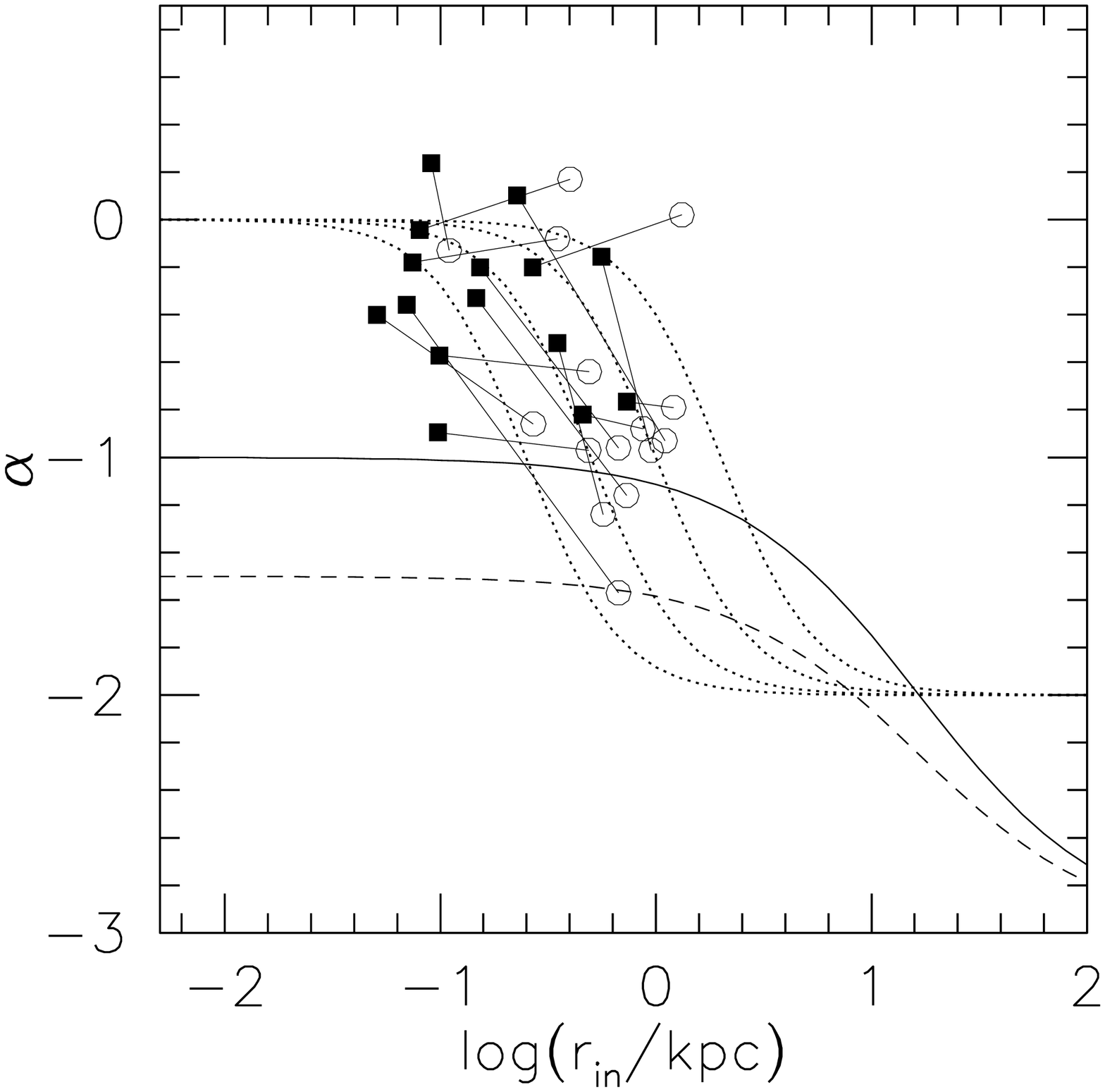}
\caption{Comparison of the inner slopes $\alpha$ and innermost
radii $r_{\rm in}$ of the high-resolution curves and the medium
resolution \HI curves shown in Fig.~\ref{hiprofiles}. The filled
squares indicate the values derived from the high-resolution curves;
open circles indicate values derived from the \HI
curves. Corresponding pairs are connected. Galaxies move along the
isothermal halo curves when resolution is increased.
\label{slopeswaters}}
\end{figure}

In Fig.~\ref{hiprofiles} we plot the inversions of the high-resolution
hybrid curves and the lower-resolution \HI curves. The slopes as
derived from these curves are shown in the Figure.  Those derived from
the \HI curves tend to be steeper than those from the equivalent
high-resolution curves.  The values of the slopes in
Fig.~\ref{hiprofiles} are slightly different from those given in
Fig.~\ref{massprofiles}. This is because the \HI profiles generally
did not extend as far inward as the hybrid profiles, and we had to
choose a larger upper fitting radius.  In Fig.~\ref{slopeswaters} we
plot the positions that the high-resolution and \HI curves occupy in
the $\alpha-r_{\rm in}$ diagram. The general trend is indeed that as
the resolution is increased, the galaxies move towards smaller slopes,
more or less along the pseudo-isothermal halo model curves.

It is interesting to compare our analysis with that of \citet{BS2001}
(hereafter vdBS), who investigated the \HI curves of a number of
galaxies for which we now have high-resolution data. Their conclusion
was that the majority of the galaxies they investigated was consistent
with a $\Lambda$CDM scenario, but they also noted that it is difficult
to distinguish between a cusp and a core model at the typical
resolution of the \HI data. It is clear that this was because the
lower spatial and velocity resolution makes it possible to force a NFW
fit on the data, even though it is not always of the shape preferred
by the data.  We will discuss the galaxies they consider in turn.

{\bf U731:} The \HI curve differs substantially from the
high-resolution curve.  Though our NFW fit gives similar parameters to
the vdBS fit ($c \sim 18$ versus their $c\sim 16$), our curve provides
a much inferior fit.  This is mainly caused by the more linear inner
curve.  The inner slope of the high-resolution mass density profile
shows a large change away from the optimum CDM value.

{\bf U3371:} The high-resolution curve is more linear and rises less
steeply than the \HI curve. The slope of the inner curve again shows a
large change: the \HI value is $\alpha \sim 1$ and consistent with
CDM, the new value is $\alpha \sim 0$ and consistent with a
constant-density core. We were unable to derive a sensible NFW model
fit.

{\bf U4325:} This galaxy shows the same behaviour as U3371: the slope
derived from the \HI data is optimum for CDM, whereas the slope
derived from the high-resolution data is consistent with a constant
density core. 

{\bf N4395 (U7524):} This is the best-resolved \HI curve in the vdBS
sample, but here we also see the same trends noted above: the slope of
the \HI curve is steeper than the slope of the high-resolution
curve. The new curve also differs markedly from the \HI curve. The
difference between NFW and pseudo-isothermal is most marked in the
inner parts. The increase in resolution has made this galaxy less
consistent with CDM.

{\bf N4455 (U7603):} vdBS note that for this galaxy no meaningful CDM
fit could be derived. We find identical slopes for both sets of data.
As a note of interest, we note that the NFW fit to the \HI curve
presented in vdBS shows the same systematic residuals as the H$\alpha$
LSB rotation curves presented in dBMR.

{\bf DDO189 (U9211):} We find similar slopes as vdBS and also the
models are of similar quality. The new data does not prefer one model
over the other.

Of the 6 galaxies in this small subsample vdBS find that 5 are
consistent with CDM, with one dubious case. The increased resolution
suggests that only one of these 6 galaxies (DDO189) is still
consistent with CDM (this is also the galaxy with the poorest
high-resolution data in the subsample). For two galaxies CDM is
perhaps inconsistent with the data (U731 and N4455), whereas for the
remaining three the data are clearly inconsistent with the CDM models.
In summary, we conclude that the slopes of rotation curves are best
described by a pseudo-isothermal halo model.

\section{Conclusions}

We have presented high-resolution $H\alpha$/\HI rotation curves of a
sample of 26 LSB and dwarf galaxies. We have fitted mass-models to
these rotation curves assuming both a pseudo-isothermal
(core-dominated) halo and a CDM NFW (cusp-dominated) halo. We find
that the pseudo-isothermal halos generally provide better fits, though
the difference is maybe not as pronounced as in dBMR, which can be
traced back to the fact that our sample contains more galaxies of
higher surface brightness than the dBMR sample.

We find more galaxies with low concentration parameters than predicted
by numerical CDM simulations. An analysis of the mass-density profiles
of the halos, as derived from minimum-disk rotation curves, shows that
the galaxies in our sample are dominated by more-or-less constant
density cores. As shown in dBMBR the few galaxies that show slopes
consistent with CDM are usually the ones that are not well-resolved,
so that one traces the edges of the constant-density cores rather than
the cores themselves. We have illustrated this explicitly by comparing
the mass-density profiles of a number of galaxies for which we have
both high-resolution $H\alpha$ curves and
medium-resolution \HI curves. In general the slopes
derived from the high-resolution curves are less steep than those from
the \HI curves.

In conclusion, our high resolution rotation data on nearby dwarfs and
LSB galaxies show that the halos of late-type LSB and dwarf galaxies
are dominated by kpc-sized constant density cores inconsistent with
the predictions of cuspy dark matter halos in cosmological numerical
simulations.

\section*{Acknowledgements}
EdB would like to thank ANSTO for their financial support which made
attending the February 2001 observing run possible. AB thanks the
Programme National Galaxies for financial support of his observing
runs at the Observatoire de Haute Provence. We thank Stacy McGaugh for
useful discussions.

\label{lastpage}

\end{document}